\documentclass[nonacm]{acmart}

\makeatletter                   
\def\mdseries@tt{m}             
\makeatother                    

\usepackage{amsmath}
\usepackage{amssymb}
\usepackage{dsfont}
\usepackage{subcaption}
\usepackage{booktabs}
\usepackage{float}
\usepackage{tabularx}
\usepackage{xspace}
\usepackage{hyperref}           
\hypersetup{
    colorlinks=true,
    linkcolor=blue,
    filecolor=red,      
    urlcolor=magenta,
    breaklinks=true,            
}
\setcopyright{none}
\DeclareRobustCommand\onedot{\futurelet\@let@token\@onedot}
\def\@onedot{\ifx\@let@token.\else.\null\fi\xspace}

\def\eg{\emph{e.g}\onedot} 
\def\ie{\emph{i.e}\onedot} 
\def\cf{\emph{c.f}\onedot} 
 
\def\wrt{w.r.t\onedot} 
\def\etal{\emph{et al}\onedot}

\newcommand{\expnumber}[2]{{#1}\mathrm{e}{#2}}

\newcommand*{\x}{{\times}}

\author{Mark Boss}
\affiliation{%
 \institution{University of T\"ubingen}
}

\author{Hendrik P.A.\ Lensch}
\affiliation{%
  \institution{University of T\"ubingen}
}

\begin{document}
\sloppy
\title{Single Image BRDF Parameter Estimation with a Conditional Adversarial Network}

\begin{teaserfigure}
  \centering
  \begin{tabular}{c @{\hskip 0.05in} c @{\hskip 0.05in} c @{\hskip 0.05in} c}
       \includegraphics[height=0.32\textwidth]{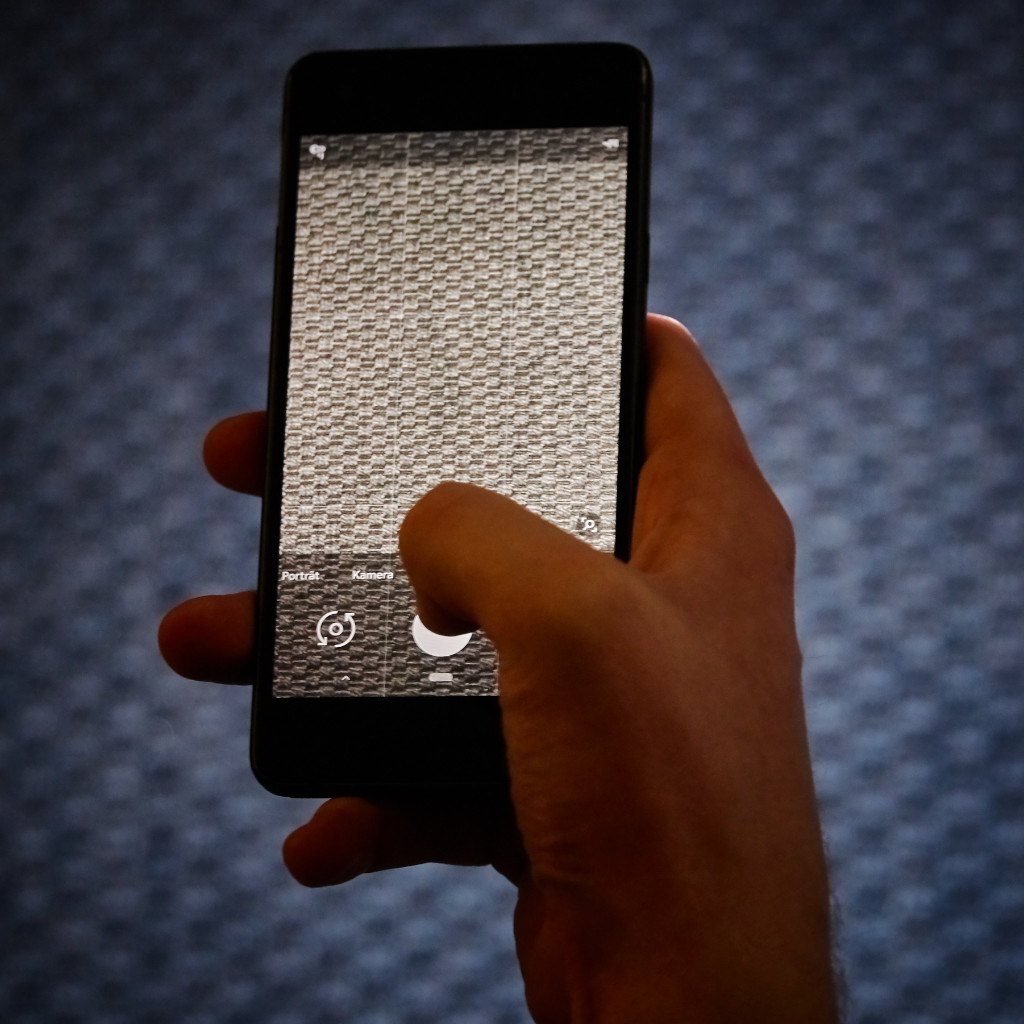} & \includegraphics[height=0.32\textwidth]{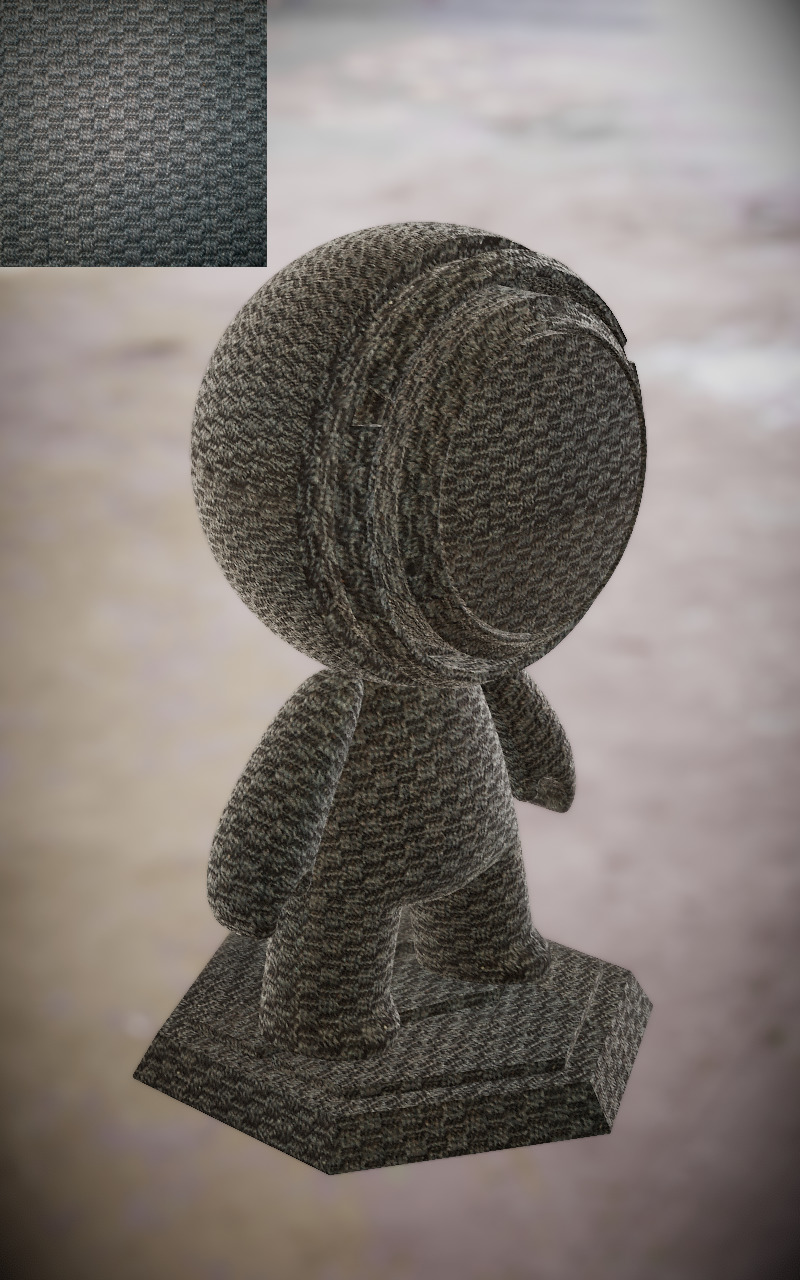} &  
       \includegraphics[height=0.32\textwidth]{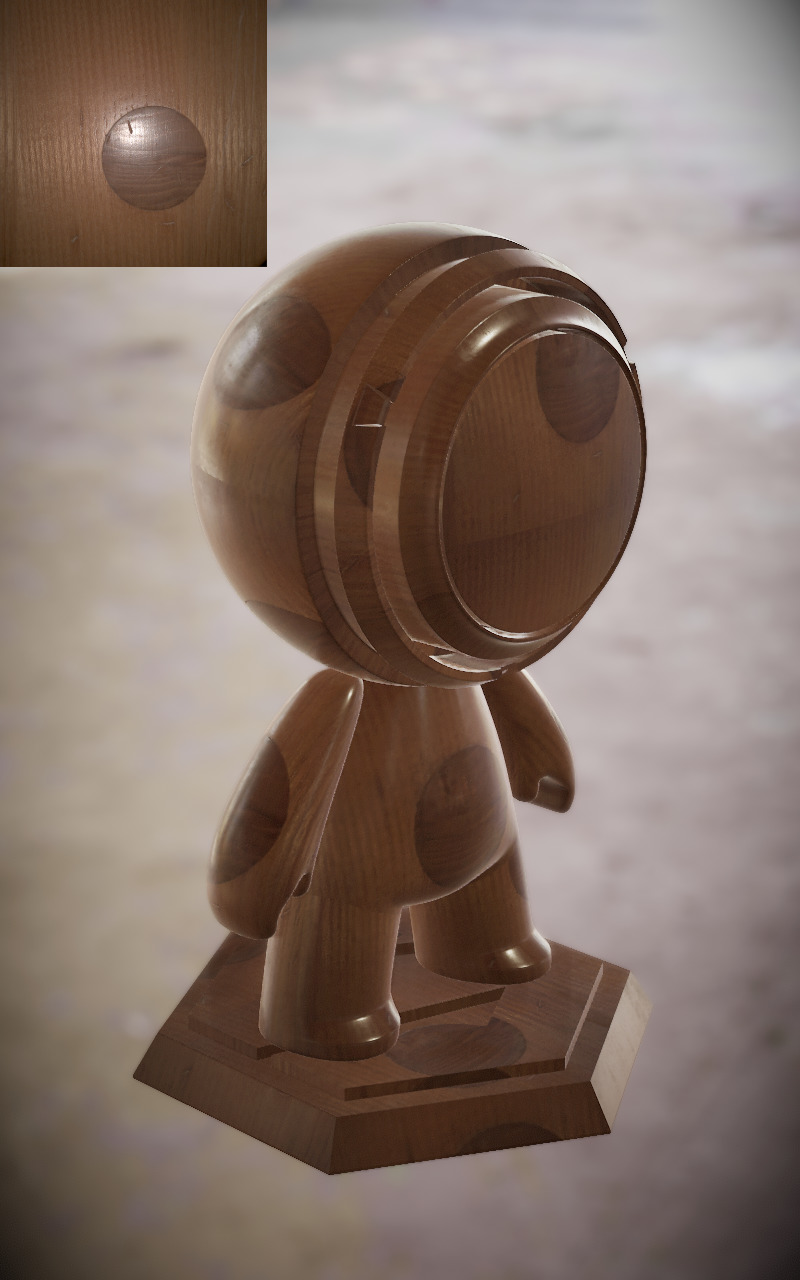}  \includegraphics[height=0.32\textwidth]{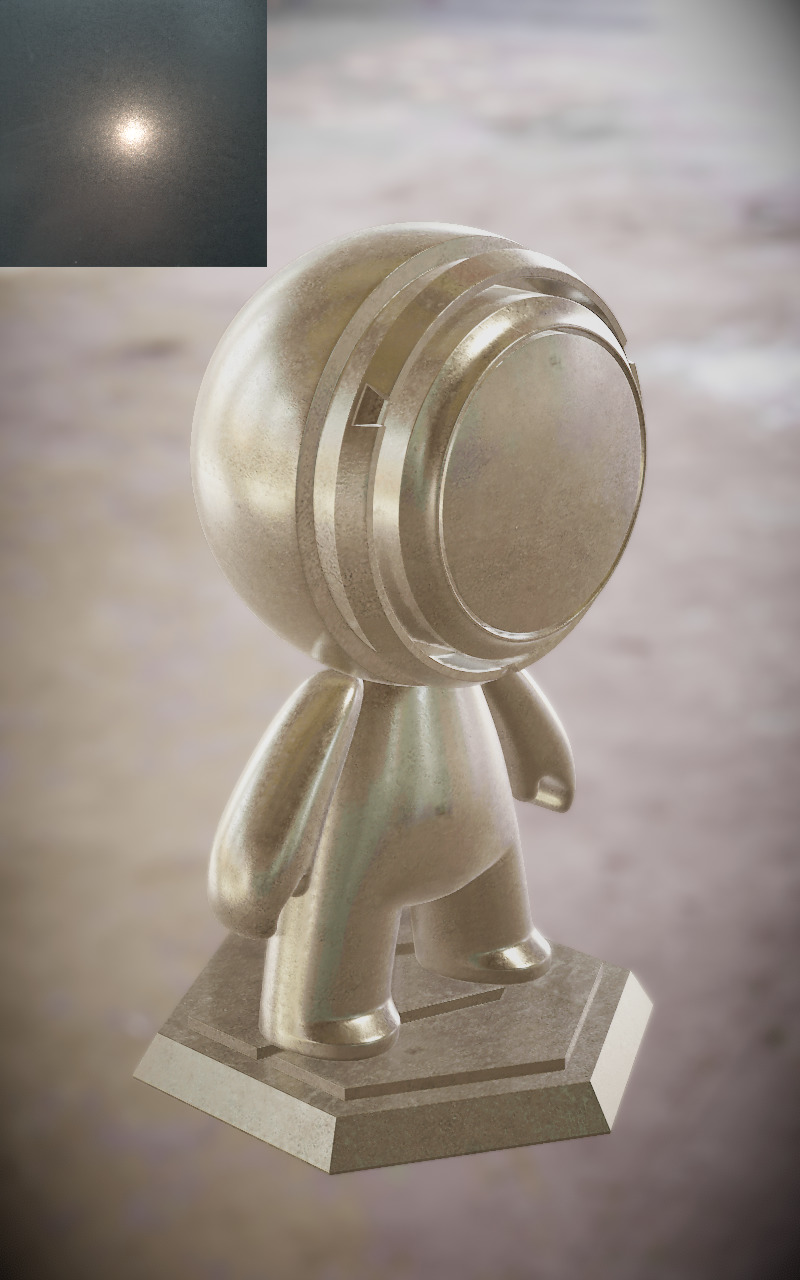} 
  \end{tabular}
  \caption{For real world surfaces our deep neural network predicts spatially-varying BRDF parameters which allow for realistic relighting. The prediction is created from a single flash photograph in a casual capture process (left most image). A selection of materials and the accompanying input photographs (upper left) are shown to the right. The 3D-Model is created by Allegorithmic and is available at: \url{https://share.allegorithmic.com/libraries/2887}.}
  \label{fig:teaser}
\end{teaserfigure}

\begin{abstract}
Creating plausible surfaces is an essential component in achieving a high degree of realism in rendering. To relieve artists, who create these surfaces in a time-consuming, manual process, automated retrieval of the spatially-varying Bidirectional Reflectance Distribution Function (SVBRDF) from a single mobile phone image is desirable. By leveraging a deep neural network, this casual capturing method can be achieved. The trained network can estimate per pixel normal, base color, metallic and roughness parameters from the Disney BRDF~\cite{Burley2012}. The input image is taken with a mobile phone lit by the camera flash. The network is trained to compensate for environment lighting and thus learned to reduce artifacts introduced by other light sources.
These losses contain a multi-scale discriminator with an additional perceptual loss, a rendering loss using a differentiable renderer, and a parameter loss. Besides the local precision, this loss formulation generates material texture maps which are globally more consistent. The network is set up as a generator network trained in an adversarial fashion to ensure that only plausible maps are produced. 
The estimated parameters not only reproduce the material faithfully in rendering but capture the style of hand-authored materials due to the more global loss terms compared to previous works without requiring additional post-pro\-ces\-sing. Both the resolution and the quality is improved.


%
\end{abstract}




\maketitle
\thispagestyle{empty}

\section{Introduction}

With the advance of processing power and improvements in rendering algorithms, movies and video games approach photorealism. Rendering algorithms recreate the behavior of light realistically on the highly detailed 3D models of characters and scenes. For a high level of realism, the correct reflectance behavior of surfaces is critical. Realistic materials are often captured using photogrammetry or with a Bidirectional Texturing Function (BTF) measurement device. Capturing materials at this quality level is often unfeasible because of budget or time constraints as either the capture process is time-consuming or the expenses to develop or acquire a measurement device is high. The other approach is that artists manually recreate these materials in software suites such as Allegorithmic Substance Designer \cite{allegorithmic}. However, achieving realistic, manually authored results is a time-consuming process. Ideally, artists want to capture materials quickly with a low-cost device such as a mobile phone.

We propose a method to generate high-quality spatially-varying Bidirectional Reflectance Distribution Function (SVBRDF) parameters from low-cost devices with a single, predominantly camera flash lit photograph of a planar surface. To further aid the artists the reconstructed BRDF parameters should not only recreate the captured material but should imitate hu\-man-au\-tho\-red materials. Reconstructing a BRDF from a single view and lighting position is highly ill-posed, and thus the main concern is to estimate BRDF parameters which mainly fulfill the human-authored material appearance. We introduce several loss terms to guide the estimation of isotropic SVBRDF parameters to enforce the human-authored style and thus solve the ambiguity of this task. The result is parameters for the popular Cook-Torrance model \cite{Cook1982} with the metallic parameter taken from the Disney BRDF \cite{Burley2012}. We estimate eight parameters per pixel from a single image input: diffuse color (3 channels), normal (3 channels), roughness (1 channel) and metallic (1 channel).

\begin{figure}[b]
    \begin{center}
    \begin{tabular}{c @{\hskip 0.07in} c @{\hskip 0.05in} c @{\hskip 0.05in} c @{\hskip 0.05in} c}
    
    Input & Diffuse & Specular & Normal & Roughness \\
    
    \raisebox{1.8\normalbaselineskip}[0pt][0pt]{\rotatebox[origin=c]{90}{Ours}} & 
    \includegraphics[width=0.12\linewidth]{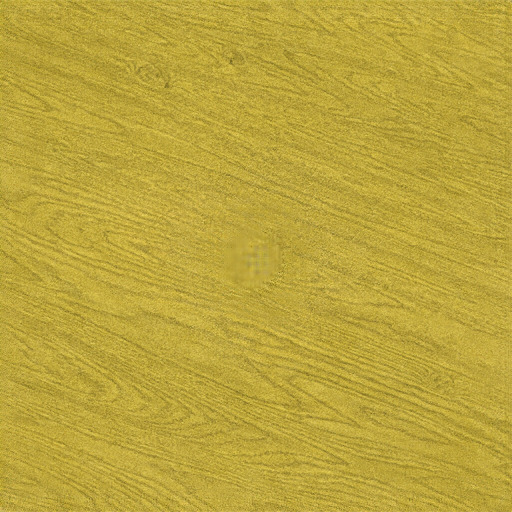} & 
    \includegraphics[width=0.12\linewidth]{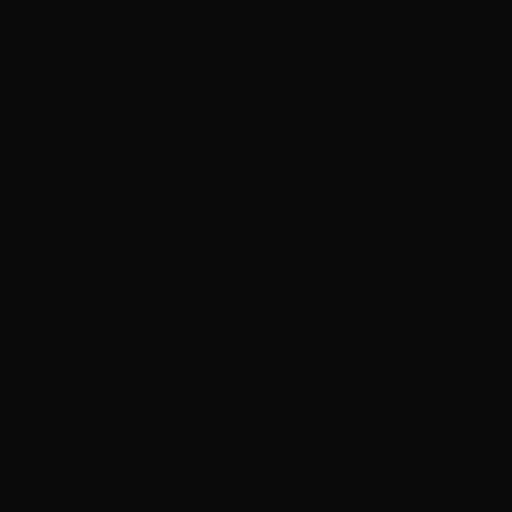} & 
    \includegraphics[width=0.12\linewidth]{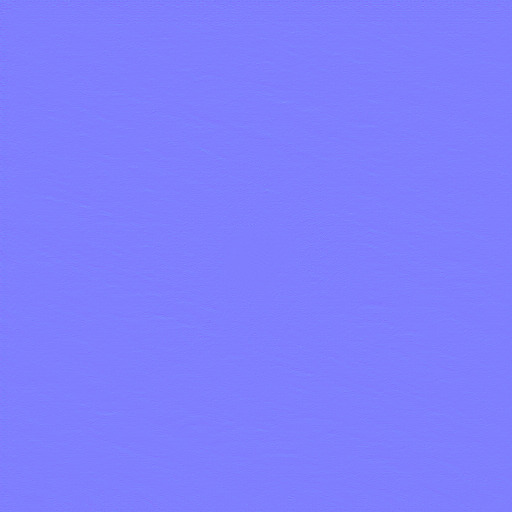} &
    \includegraphics[width=0.12\linewidth]{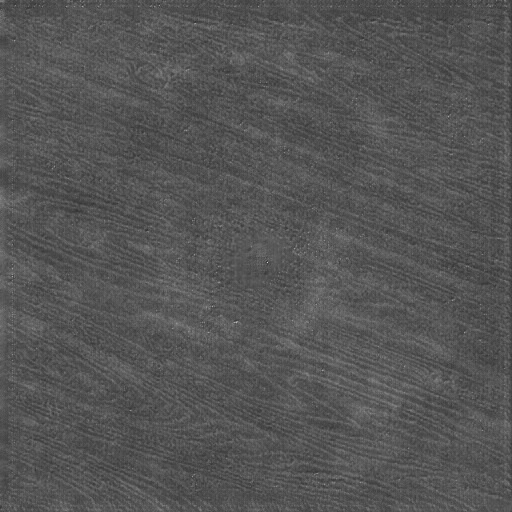} \\
    
    \includegraphics[width=0.12\linewidth]{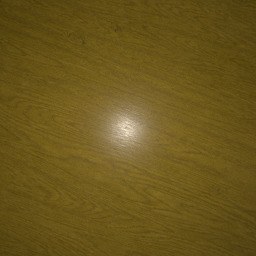} & 
    \includegraphics[width=0.12\linewidth]{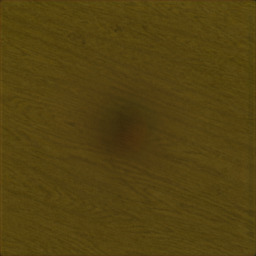} & 
    \includegraphics[width=0.12\linewidth]{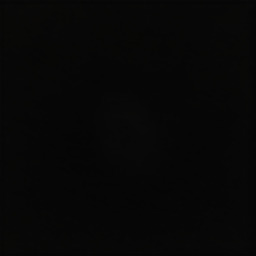} & 
    \includegraphics[width=0.12\linewidth]{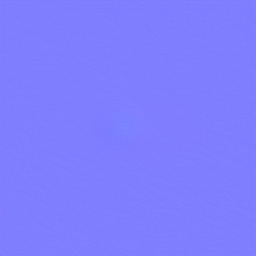} &
    \includegraphics[width=0.12\linewidth]{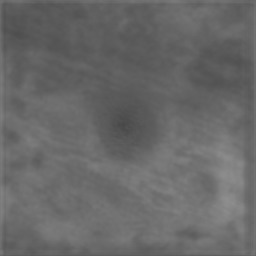} \\
    
    \\
    
    \includegraphics[width=0.12\linewidth]{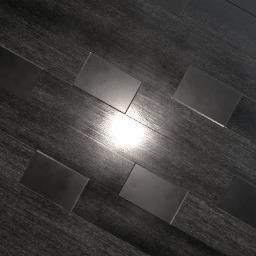} & 
    \includegraphics[width=0.12\linewidth]{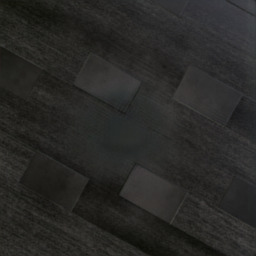} & 
    \includegraphics[width=0.12\linewidth]{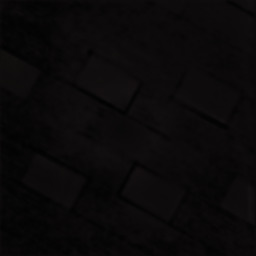} & 
    \includegraphics[width=0.12\linewidth]{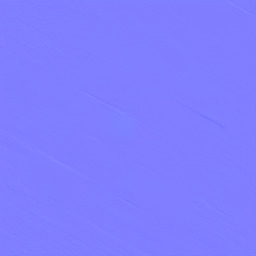} &
    \includegraphics[width=0.12\linewidth]{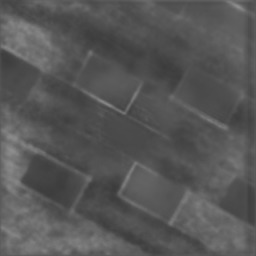} \\
    
    \raisebox{1.8\normalbaselineskip}[0pt][0pt]{\rotatebox[origin=c]{90}{Ours}} & 
    \includegraphics[width=0.12\linewidth]{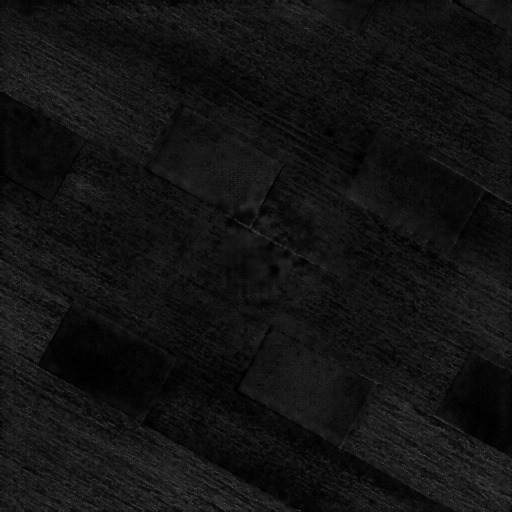} & 
    \includegraphics[width=0.12\linewidth]{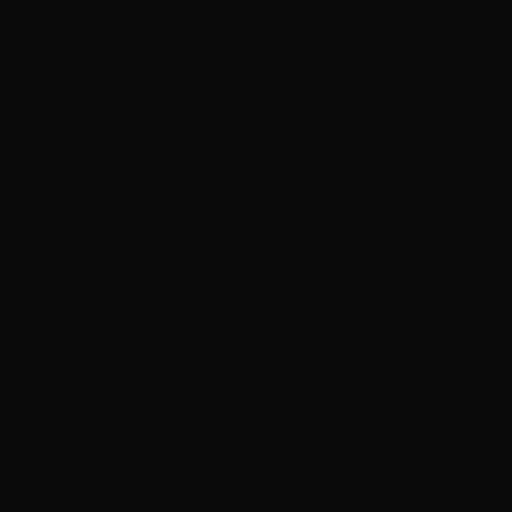} & 
    \includegraphics[width=0.12\linewidth]{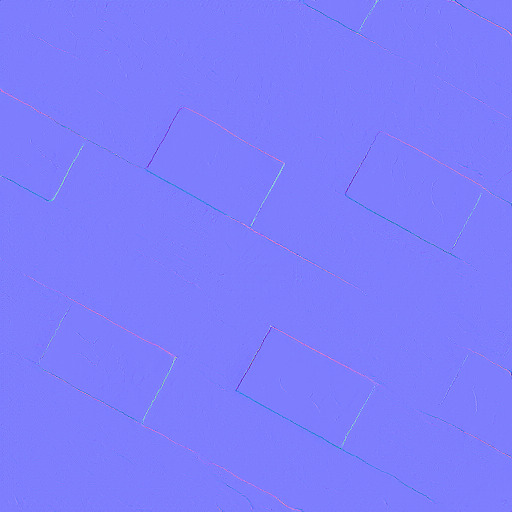} &
    \includegraphics[width=0.12\linewidth]{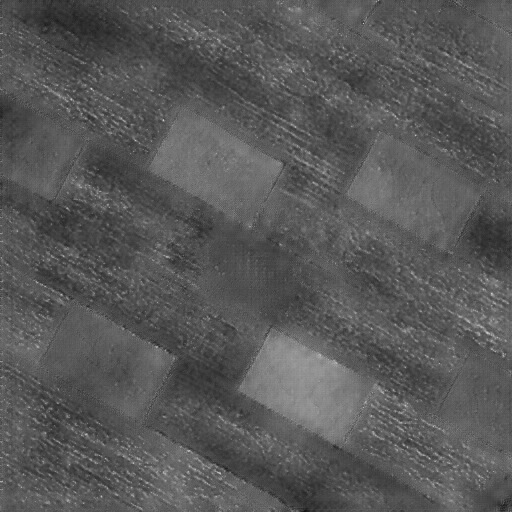} \\
    
    \end{tabular}

    \end{center}
    \caption{Showcase of several current problems in single shot BRDF estimation. We compare the prediction of our method to a previous method \cite{Deschaintre2018}. Notice the artifact by the harsh specular camera flash of the top material in the diffuse and roughness map of the previous method, which is not or hardly visible in our result. In the bottom example, the previous method did not remove secondary illumination. For example, the diffuse and roughness parameter map displays the light fall off and the secondary illumination in the top right corner. We remove these artifacts in our prediction (bottom).}
    \label{fig:problem_examples}
\end{figure}

Compared to previous work which requires additional post-pro\-ces\-sing \cite{Li2018} or added a priori knowledge \cite{Deschaintre2018} to reduce artifacts from harsh specular reflections, our method reduces these artifacts through our novel loss formulation. Fig. \ref{fig:problem_examples} visualizes several of these problems. This loss formulation forces the generator to learn how to reduce these artifacts, which results in an overall improved prediction.
The key contributions of this work are: 
\begin{itemize}
    \item We introduce a Generative Adversarial Network (GAN) architecture with multi-scale discriminators to reduce artifacts from specular highlights. The loss focuses more on global structure consistency, and the resulting parameter maps incorporate the human-authored appearance more closely compared to previous works. This method proofs that a perceptually based loss is helpful in BRDF parameter estimation.
    \item An increase in the SVBRDF parameter map resolution of a factor of four results in greater detail compared to previous work. Detail is an essential aspect of high-quality BRDFs.
    \item A large procedurally generated BRDF training dataset rendered under varying environment illumination to capture realistic recording scenarios with a mobile phone. The dataset contains 40544 materials with three times different environment illumination.
\end{itemize}

\section{Related Work}

The concept of capturing BRDFs with an extremely sparse measurement from a single image is an active area of research. Several methods address the ill-posedness of the problem.

\paragraph{Optimization-based Planar Surface BRDF Estimation} \citet{Aittala2015} use the property that materials are often stationary in an optimization approach to fit a single tile of an image to other tiles. The result is an SVBRDF from a small low-resolution area of the material. In a follow-up work, \citet{Aittala2018} refine the tile-based optimization approach by using a pre-trained neural network as a perceptual loss in the optimization and additionally add a differentiable renderer in the optimization.

\paragraph{Deep Neural Network Based Planar Surface BRDF Estimation} \citet{Li2017} are the first, who explore the possibility of using Convolutional Neural Networks (CNN) for BRDF estimation from single input images. The result is spatially-varying information about the diffuse and normal parameters and homogeneous parameters for the roughness and specular information. Recently, \citet{Li2018} and \citet{Deschaintre2018} proposed new network architectures to improve the prediction quality. Both methods introduced a rendering loss where the parameters are evaluated with a differentiable renderer to provide the network with additional information about other lighting conditions. To further improve the result \citet{Li2018} added a classification network, which provides additional information about the rough material category to be used in the decoding step of an encoder-decoder network. They further use a post-processing step based on Conditional Random Fields to enhance the quality of each parameter map. \citet{Deschaintre2018} introduce a global feature track which extracts mean feature vectors by pooling in each encoding and decoding step. They then add these feature vectors to the encoder-decoder network to extract non-local information to aid reducing specular artifacts from the harsh flashlight of the captured surfaces.

\paragraph{BRDF Estimation of 3D Objects} Another novel topic is the joint estimation of shape and appearance of 3D objects in a single photograph. For this complex task, the authors need to estimate illumination, shape, and material, which all interact with each other generating complex ambiguities. Recently, two methods tackle this issue. \citet{Nam2018} use several unstructured flash-lit photographs of an object. An optimization approach starts with an initial point cloud estimation using Structure from Motion (SfM) and continues with detailed surface normals and the appearance. The process then uses the lower ambiguity from the now known reflectance properties to refine the geometry. \citet{Li2018a} estimate appearance, shape, and illumination from a single predominately flash lit photograph. To achieve this, they use a deep neural network to learn all tasks jointly. In a first step, they estimate an initial result for albedo, normal, roughness, depth and the environment map represented by spherical harmonics. They then pass the result to a differentiable renderer which renders the direct light and environment map approximation. A global light approximation using a pre-trained network adds the contribution of the first three light bounces and the image is used to calculate the loss against the input image. To further improve the resulting parameters they use two additional cascaded networks, which refine the parameters iteratively.

\paragraph{Style Transfer and Perceptual Losses} Lastly, the work on image-to-image translation is an important area, as BRDF estimation from a single image can be seen as a style transformation, \ie from an illuminated surface to its reflectance parameters. One of the first general frameworks for these tasks is proposed by \citet{Isola2017}, which can handle a wide variety of translations. They base it on a Generative Adversarial Network (GAN) architecture. Recently, perceptual losses gained traction in tasks like super resolution and style transfer and surpassed state of the art using pre-trained networks for extraction of features and matching these features from prediction to ground truth \cite{Johnson2016, enhancenet, wang2018pix2pixHD, Dosovitskiy2016, Gatys2016ImageST}.

\section{Material Representation}
\label{sec:material_representation}

To support artists in material creation, our framework estimates parameters for a popular BRDF model often used in modern games such as Unity\footnote{\url{https://unity3d.com}} or the Unreal Engine\footnote{\url{https://unrealengine.com}}: the Cook Torrance model with the metallic term introduced by Burley in the Disney BRDF \cite{Burley2012}. The BRDF is described as: $f_r = k_s + k_d$. This model consists of the specular lobe $k_s$ and the diffuse lobe $k_d$. The specular lobe of the Cook Torrance model is defined as:
\begin{equation}
    k_{\text{s}} = \frac{D(\alpha, \omega_i, \omega_o, n) F(\omega_i, \omega_o) G(\alpha, \omega_i, \omega_o, n)}{4(\omega_o \cdot n)(n \cdot \omega_i)},
    \label{eq:cook_torrance}
\end{equation}

where $D$ is the normal distribution function, $F$ is the Fresnel term, $G$ is the geometric attenuation factor, $\alpha$ is the roughness parameter, $n$ is the surface normal, and $\omega_i$, $\omega_o$ being the incoming and outgoing direction. For the diffuse lobe $k_d$ the Disney term is used \cite{Burley2012}. The GGX normal distribution function drives the microfacet model.

As the metallic term only requires a single color map, the base color, the specular and diffuse color needs to be extracted with the help of the metallic map. Here, an assumption is made. For non-metallic surfaces, the specular color is assumed to be $0.04$. This means a 4\% base reflectance. The diffuse color $d$ can then be calculated using $d = b (1 - m)$ and the specular color $s$ using $s = 0.04 (1 - m) + b m$, with $b$ being the base color and $m$ being the metallic value. A metallic value of $1$ means metallic and $0$ means non-metallic. The base color map needs to be transformed from sRGB to linear color space for this operation. One large advantage of using this model is that fewer parameters need to be estimated. For the base color and metallic case, only four parameters are required while for specular and diffuse color six parameters are to be predicted.

\section{Network Architecture and Loss Formulation}

\begin{figure*}[hbt!]
  \centering
  \includegraphics[width=0.9\textwidth]{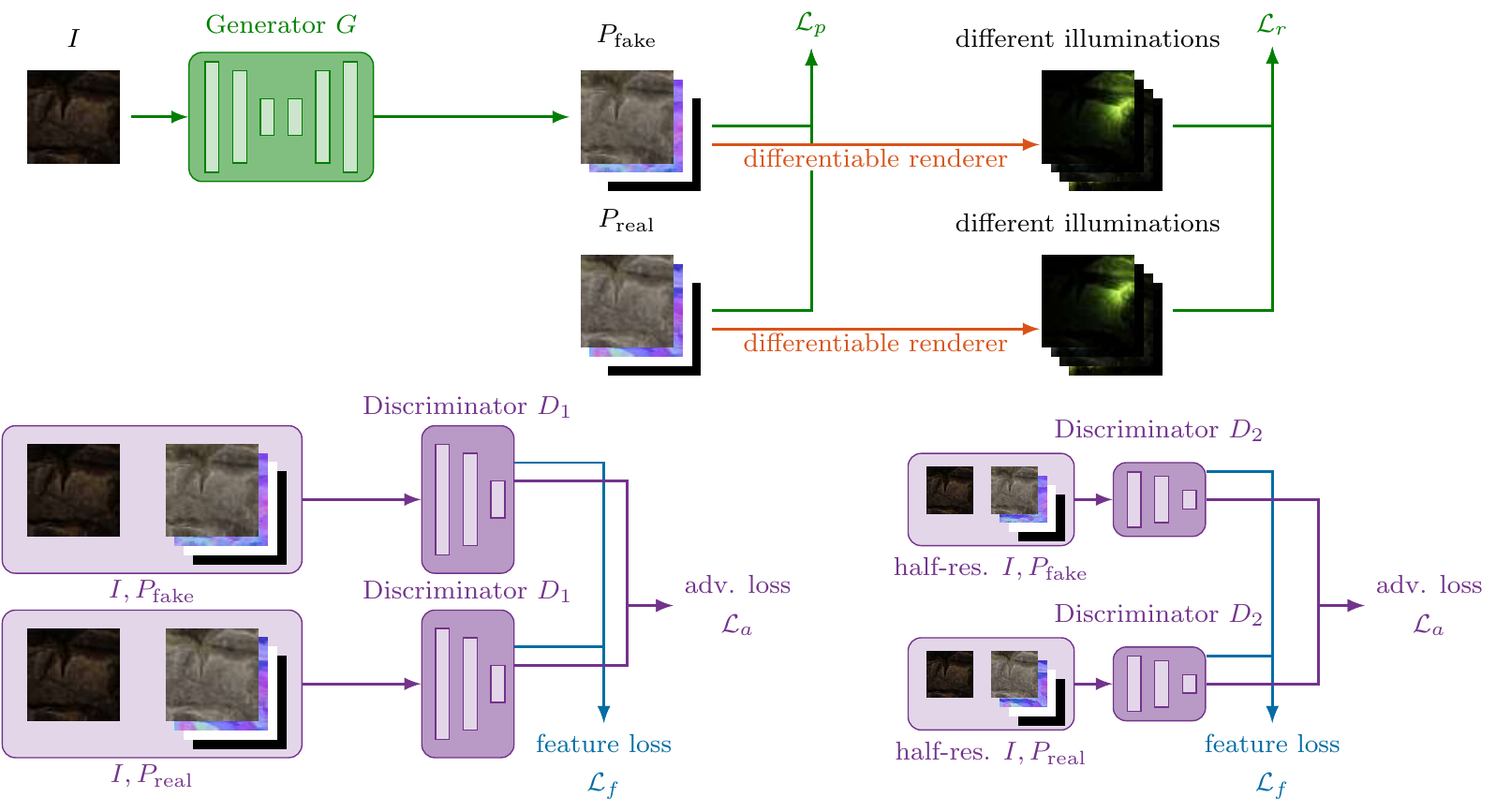}
 \captionof{figure}{Conditional generative adversarial network with rendering and multi-scale discriminator loss. The generator network and its input $I$ as well as the output $P_\text{fake}$ are highlighted in green. The content loss compares the generated parameters $P_\text{fake}$ and the ground truth parameters $P_\text{real}$ as well as their respective rendered loss images. Here, $\mathcal{L}_p$ calculates the MAE loss on the parameter images and the rendering loss $\mathcal{L}_r$ on the rendered parameters using a differentiable renderer. The discriminators $D_1$ and $D_2$ with their inputs and loss terms are highlighted in purple. The loss terms include an adversarial $\mathcal{L}_a$ and feature loss $\mathcal{L}_f$. It is worth noting that $D_2$ receives the same input as $D_1$ but in half the resolution.}
 \label{fig:overview}
\end{figure*}

Achieving a high prediction resolution and at the same time plausible BRDF parameters, which capture the look and feel of human-authored materials, several challenges need to be addressed.

Common loss terms such as a loss against the ground truth parameters or a rendering loss~\cite{Aittala2018, Li2018a, Deschaintre2018, Nam2018, Li2018} provide a reliable per-pixel based training signal. Due to this locality, even with the advanced rendering loss using a differentiable renderer, the BRDF is often not reliably estimated under every viewing and lighting direction. 
To capture plausible parameters matching the style of human-authored materials for the entire texture an adequate style loss is required, which is defined on larger areas than the common per-pixel losses. We propose a GAN architecture with two discriminators to learn the style of human-authored materials. Based on PatchGAN~\cite{Isola2017} the first discriminator ($D_1$) receives the input in full resolution, whereas the second discriminator ($D_2$) receives the input in half resolution. This way the first discriminator $D_1$ is responsible for detecting fine details and the second $D_2$ for larger features. An overview of this structure is visualized in Figure~\ref{fig:overview}. We sum up the losses from both scales.
Importantly, the discriminators conditionally perform their fake or ground-truth prediction given the input photo $I$ and the BRDF parameter $P$.
Thus, the discriminator learns whether the parameters are plausible for a given input photograph and not if the parameters are from the material training corpus.

Note that during the inference, the prediction of BRDF parameters from an unseen image is purely based on the generator. 

The loss terms serve the primary purpose of training the parameters of the generator taking care of all the mentioned concepts.
As the U-Net~\cite{Ronneberger2015} tends to become unstable at higher resolutions the generator architecture is based on the generator of \citet{Johnson2016}. The structure of this network and the two discriminators are described in greater detail in Section~\ref{subsec:network_architecture}.

\subsection{Loss Formulation}
\label{subsec:loss_formulation}

Overall, our formulation combines three losses:
\begin{itemize}
    \item \textbf{The adversarial loss} ensures that the output parameters are in general behave similar to other BRDF parameters in the data set. This is achieved in conditional generative adversarial training while fitting the predicted parameters to the input image. 
    \item \textbf{The feature loss} stabilizes the training procedure by minimizing the distance of high-level features between ground truth and predictions in the discriminator network. 
    \item \textbf{The content loss} is based on a \textbf{parameter loss} and a \textbf{rendering loss}. The parameter loss evaluates the predicted parameter maps directly against the ground truth to provide additional information about ambiguous features in the input image. As there exist ambiguities, the rendering loss enforces plausible material parameter predictions when evaluated against the ground truth under different illumination conditions in the image domain rather than in the parameter domain. 
\end{itemize}
These loss terms are then combined to a total loss $\mathcal{L}_t$ for the discriminator and generator. 


\paragraph{Content Loss} The content loss is a sum of two different losses. The first one is the parameter loss $\mathcal{L}_p$. Here, the generated BRDF parameters from the generator are compared to the ground truth parameters. For every parameter map except the surface normals the Mean-Absolute-Error (MAE): $\ell_1(A,B)=\frac{1}{n} \sum_{i=1}^n \left| A_i - B_i\right|$ is used,, where $A$ and $B$ are the parameter maps with their $n$ elements. The $\ell_1$ loss is preferred over of the Mean-Squared-Error (MSE) loss, as it tends to produce sharper details in the predictions~\cite{Zhao2017}. Intuitively, the MSE loss tends to punish larger errors more than smaller ones, but details are more likely visible in small value changes. The linear behavior of the MAE loss enforces these small details.

As the normal map encodes a normalized vector pointing into the direction of the surface normal, a different error metric is used. The angular distance is comparable to the MAE loss in its linear behavior. It is element-wise defined as %
\begin{equation}
    \ell_\sphericalangle(a, b) = \frac{\cos^{-1} \left( a \cdot b \right)}{\pi},
\end{equation}
with $\cdot$ being the dot product between two normalized vectors $a, b$ from given normal maps.

To capture the effect of different parameters maps under various illumination conditions, a differentiable renderer produces ten re-renderings with randomly sampled viewing and lighting directions from the upper hemisphere given the ground truth and the generated parameter maps.
Five of these random light and view positions are chosen such that they contain a specular highlight by mirroring one randomly sampled direction on a random surface point. The rendering loss $\mathcal{L}_r$ captures various effects, \eg, by changing the light color randomly metallic reflections are explored. To suppress the large value range in the re-renderings with point light sources, the re-renderings $x$ are monotonously transformed by $\log(1+x)$. Instead of concentrating only on the error in the specular highlights, this step emphasizes the influence of all parameters. 
The MAE loss function is used to calculate the difference in the re-renderings yielding $\mathcal{L}_r$.

\paragraph{Adversarial Loss} An adversarial loss is utilized to enforce parameter map predictions which are indistinguishable from those in the training set. Stability in GAN training is often a problem. Empirical evidence suggests that the LSGAN loss by \citet{Mao2017} improves stability compared to the classical cross-entropy based GAN loss~\cite{goodfellow2014generative}. The LSGAN loss is defined as: %
\begin{equation}
    \min_{D} \mathcal{L}_{a}(D)  = \frac{1}{2} \mathds{E}_{x\sim p_{\text{data}}(x)} [(D(x) - 1)^2]
    + \frac{1}{2} \mathds{E}_{z\sim p_z(z)}[D(G(z))^2].
\end{equation}
The term $\mathds{E}_{x\sim p_{\text{data}}(x)} [(D(x) - 1)^2]$ forces the discriminator to classify real samples $x$ as $1$ and $\mathds{E}_{z\sim p_z(z)}[D(G(z))^2]$ ensures fake samples $G(z)$ being classified as $0$ by the discriminator. In
\begin{equation}
    \min_{G} \mathcal{L}_{a}(G)  = \frac{1}{2} \mathds{E}_{z\sim p_z(z)}[D(G(z)) - 1)^2]
\end{equation}
the generator on the other hand tries to generate samples $G(z)$ which fool the discriminator into classifying them as $1$, eventually synthesizing realistic parameter maps from input images $z$.

The discriminator input is a combination of the parameter map and one illuminated image of the material. This way, the conditional adversarial approach learns whether the predicted BRDF parameters are matching the input image rather than just focusing on some features in the parameter set. 

\paragraph{Feature Matching Loss} The structure of the discriminator acts to some extent as an encoder on these samples, \ie it is focusing on and extracting the most important features in the samples. To further improve the adversarial loss $\mathcal{L}_a$ an additional feature matching loss between the first four layers from each of the two discriminators $D_1, D_2$ is calculated. More specifically, the difference in the feature outputs of each of these layers between the real classification $D(s,p)$ and the fake classification $D(s,G(s))$ are calculated. Here, $s$ is defined as the input rendered image, $p$ as the ground truth parameters and $G(s)$ as the generated fake parameters. The loss is then calculated as:
\begin{equation}
    \mathcal{L}_f = \frac{1}{S} \sum_{k=1}^S \frac{1}{L} \sum_{i=1}^L \frac{1}{N} [D_k^i(s,p) - D_k^i(s, G(s))]^2,
\end{equation}
where $S$ is the number of discriminator scales, $L$ is the number of layers which are compared and $N$ is the number of outputs in each layer.

\paragraph{Total Loss}
For the generator, the total loss functions consist of the content loss, feature matching, and adversarial loss. It is calculated as:
\begin{equation}
    \min_{G} \mathcal{L}_t(G) = \frac{1}{4} (\mathcal{L}_a(G) + \mathcal{L}_f + \mathcal{L}_p + \mathcal{L}_r).
\end{equation}

The final discriminator loss is a combination of the adversarial loss and the feature matching loss
\begin{equation}
    \min_{D} \mathcal{L}_t(D) = \mathcal{L}_a(D) + \lambda\mathcal{L}_f
\end{equation}
with $\lambda=0.01$.

\subsection{Ablation Study}

\begin{table*}[htb!]
    \centering
\begin{tabular}{@{}l|l|ll|ll|ll|ll@{}}
\toprule
Map       & Proposed      & $-\mathcal{L}_r$ & \% worse                  & $-\mathcal{L}_p$ & \% worse                  & $-\mathcal{L}_f$  & \% worse               & $-\mathcal{L}_a$ & \% worse                      \\ \midrule
Diffuse   & 0.059         & 0.066            & {\color[HTML]{FE0000} -11.74} & 0.066  & {\color[HTML]{FE0000} -11.83} & 0.064 & {\color[HTML]{FE0000} -8.16} & 0.066       & {\color[HTML]{FE0000} -11.75} \\
Specular  & 0.047         & 0.052            & {\color[HTML]{FE0000} -10.93} & 0.052  & {\color[HTML]{FE0000} -12.64} & 0.050 & {\color[HTML]{FE0000} -7.14} & 0.052      & {\color[HTML]{FE0000} -12.41} \\
Normal    & 0.094         & 0.101            & {\color[HTML]{FE0000} -7.44}  & 0.095  & {\color[HTML]{FE0000} -1.24}  & 0.099 & {\color[HTML]{FE0000} -4.90} & 0.104       & {\color[HTML]{FE0000} -10.21} \\
Roughness & 0.111         & 0.117            & {\color[HTML]{FE0000} -5.99}  & 0.119  & {\color[HTML]{FE0000} -7.27}  & 0.120 & {\color[HTML]{FE0000} -8.44} & 0.114      & {\color[HTML]{FE0000} -3.17}  \\ \bottomrule
\end{tabular}

    \caption{Mean error over the test dataset of 7175 samples for various disabled loss terms. Each term is important.}
    \label{tab:ablation}
\end{table*}

\begin{figure}
    \begin{center}
\begin{tabular}{p{1mm} c @{\hskip 0.05in} c @{\hskip 0.05in} c @{\hskip 0.05in} c}

 & Base color & Normal & Roughness & Metallic \\

\raisebox{1.8\normalbaselineskip}[0pt][0pt]{\rotatebox[origin=c]{90}{GT}} &
\includegraphics[width=0.11\linewidth]{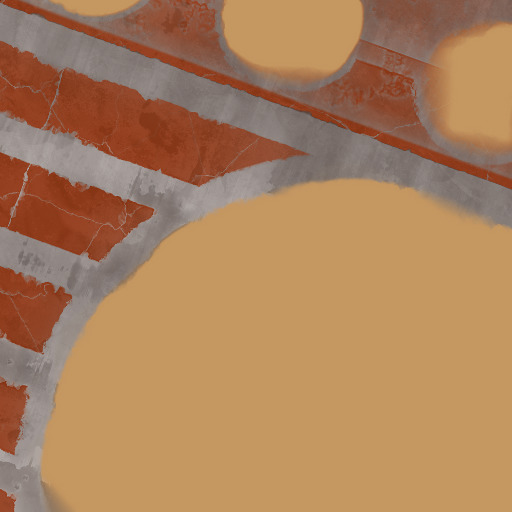} &
\includegraphics[width=0.11\linewidth]{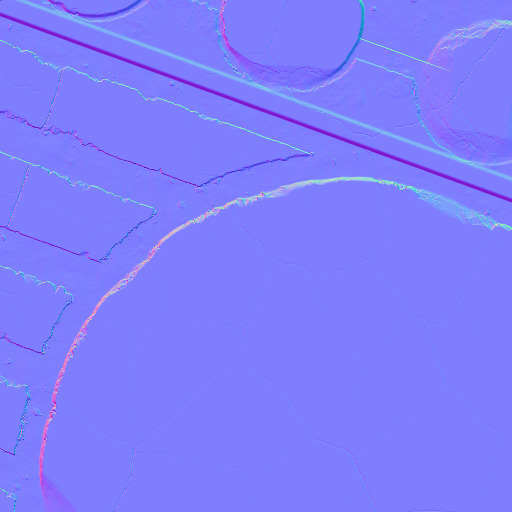} &
\includegraphics[width=0.11\linewidth]{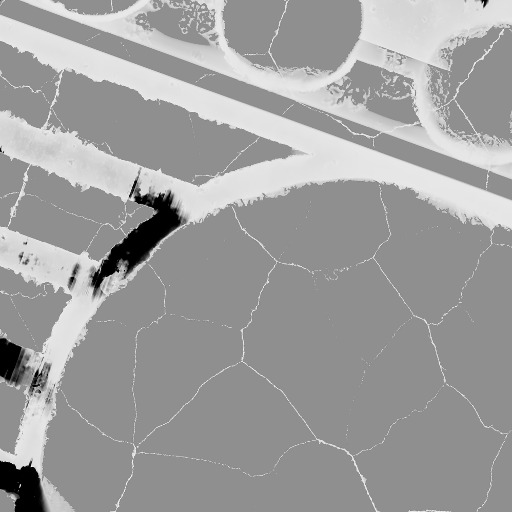} &
\includegraphics[width=0.11\linewidth]{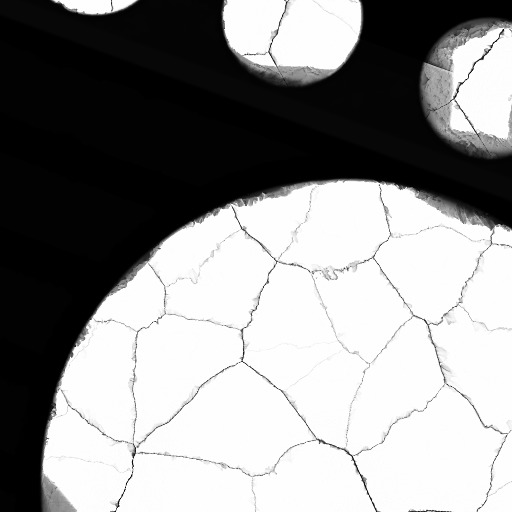} \\

\raisebox{1.8\normalbaselineskip}[0pt][0pt]{\rotatebox[origin=c]{90}{Proposed}} &
\includegraphics[width=0.11\linewidth]{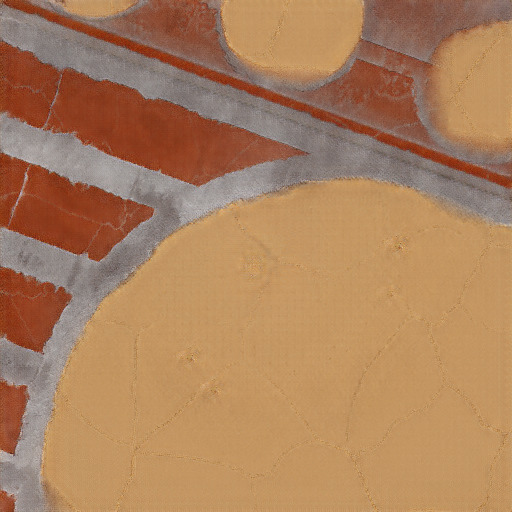} &
\includegraphics[width=0.11\linewidth]{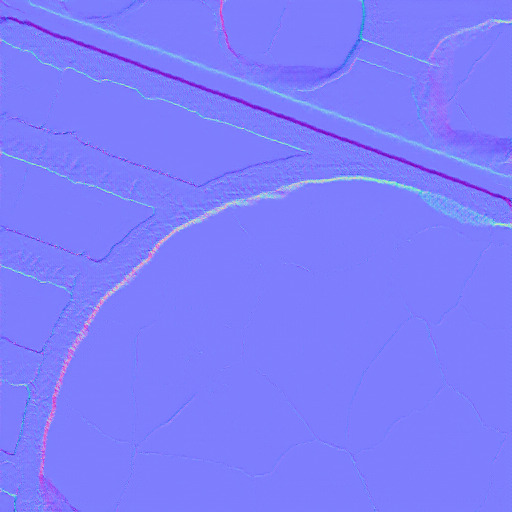} &
\includegraphics[width=0.11\linewidth]{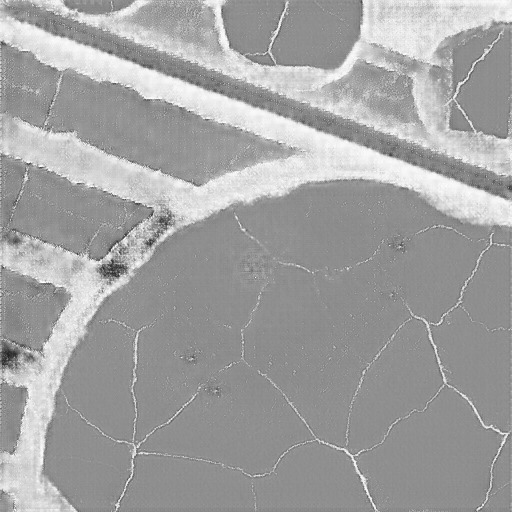} &
\includegraphics[width=0.11\linewidth]{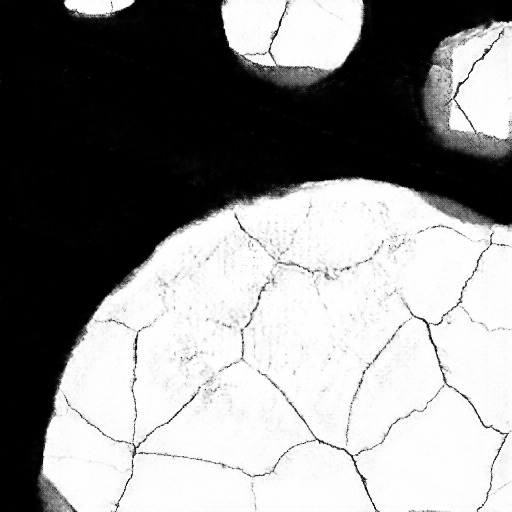} \\

\raisebox{1.8\normalbaselineskip}[0pt][0pt]{\rotatebox[origin=c]{90}{-$\mathcal{L}_r$}} &
\includegraphics[width=0.11\linewidth]{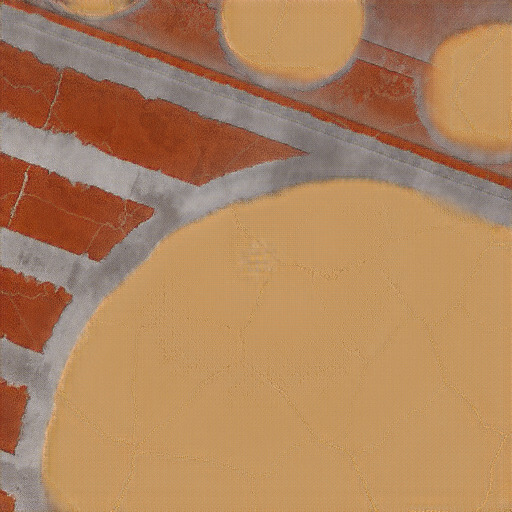} &
\includegraphics[width=0.11\linewidth]{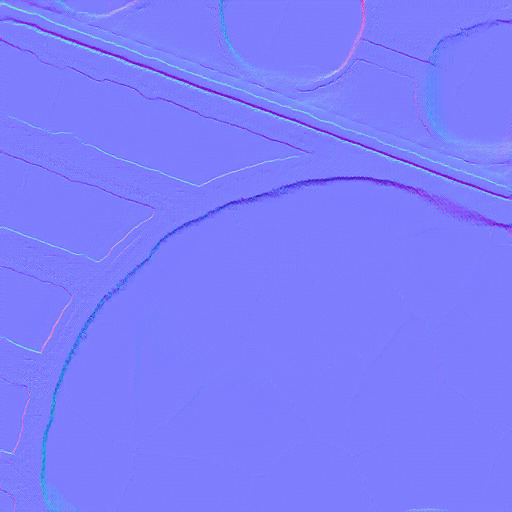} &
\includegraphics[width=0.11\linewidth]{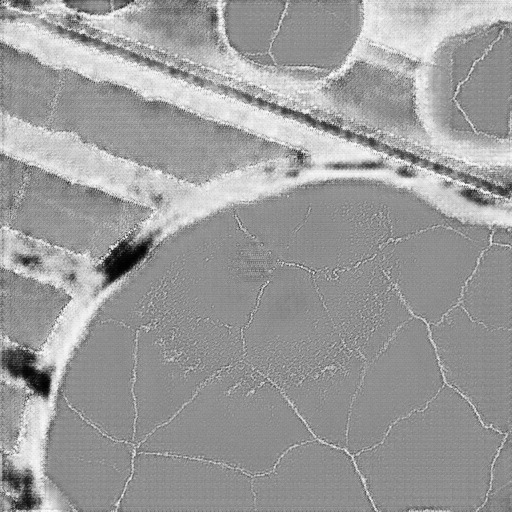} &
\includegraphics[width=0.11\linewidth]{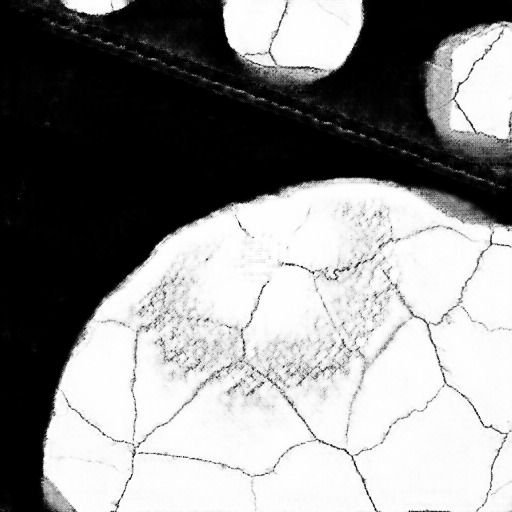} \\

\raisebox{1.8\normalbaselineskip}[0pt][0pt]{\rotatebox[origin=c]{90}{-$\mathcal{L}_p$}} &
\includegraphics[width=0.11\linewidth]{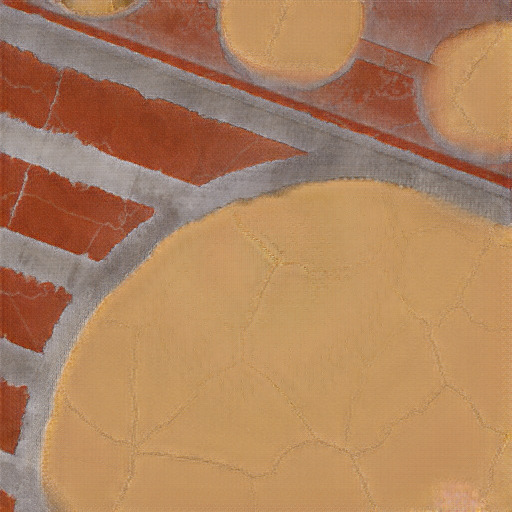} &
\includegraphics[width=0.11\linewidth]{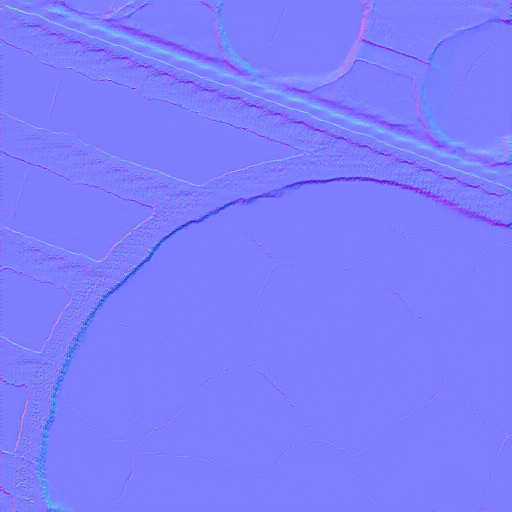} &
\includegraphics[width=0.11\linewidth]{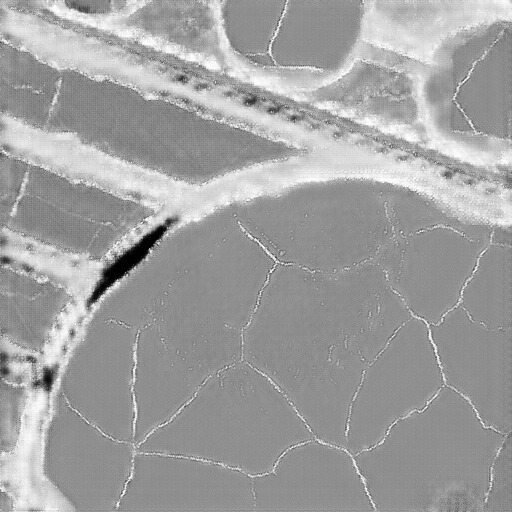} &
\includegraphics[width=0.11\linewidth]{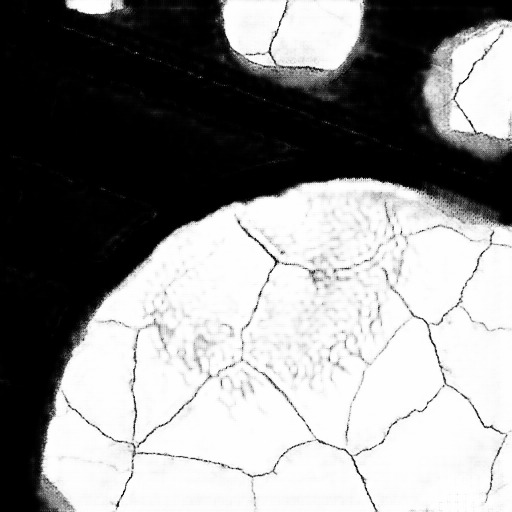} \\

\raisebox{1.8\normalbaselineskip}[0pt][0pt]{\rotatebox[origin=c]{90}{-$\mathcal{L}_a$}} & 
\includegraphics[width=0.11\linewidth]{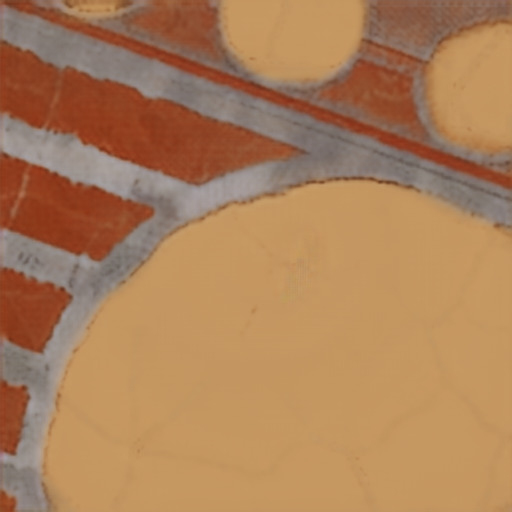} &
\includegraphics[width=0.11\linewidth]{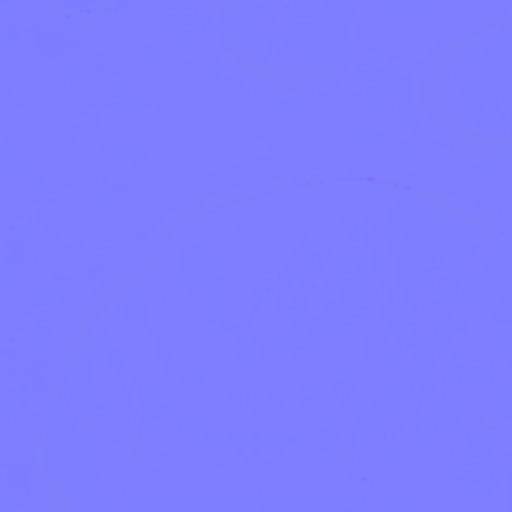} &
\includegraphics[width=0.11\linewidth]{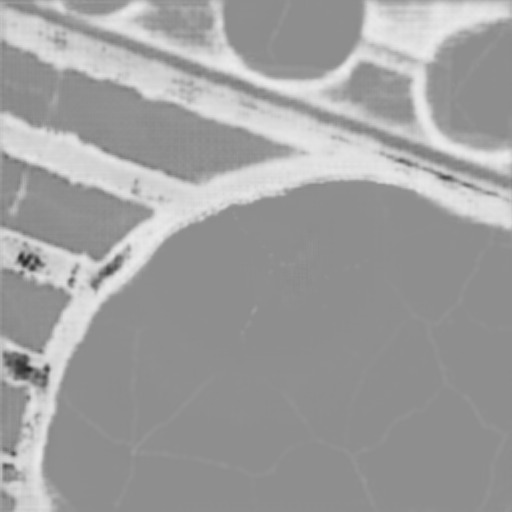} &
\includegraphics[width=0.11\linewidth]{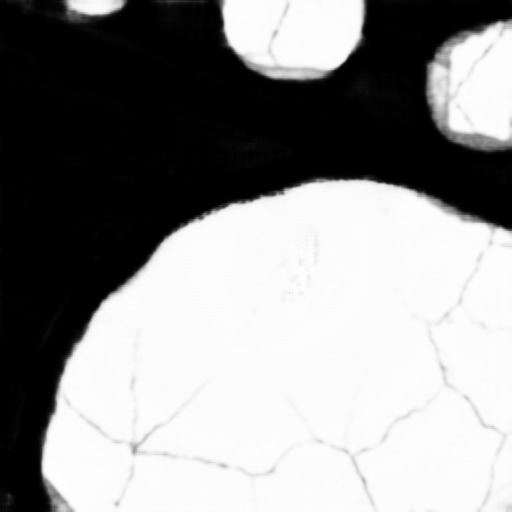} \\

\raisebox{1.8\normalbaselineskip}[0pt][0pt]{\rotatebox[origin=c]{90}{-$\mathcal{L}_f$}} & 
\includegraphics[width=0.11\linewidth]{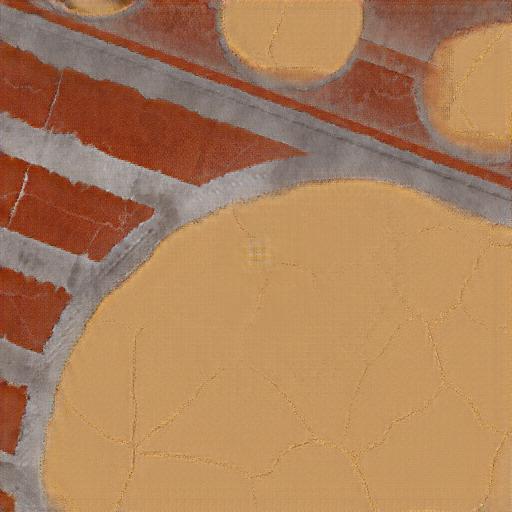} &
\includegraphics[width=0.11\linewidth]{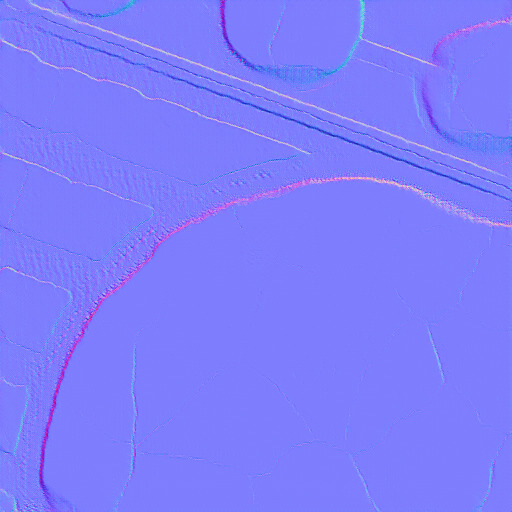} &
\includegraphics[width=0.11\linewidth]{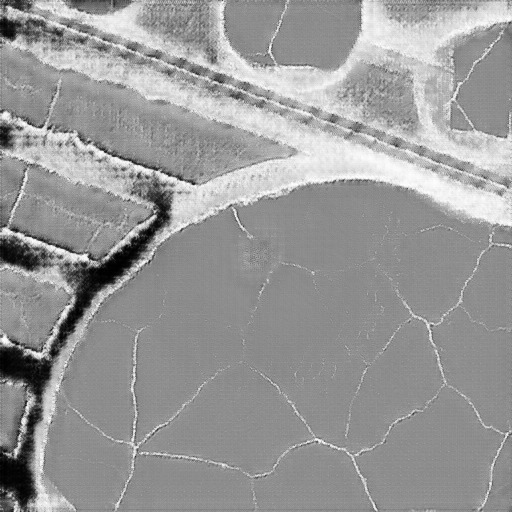} &
\includegraphics[width=0.11\linewidth]{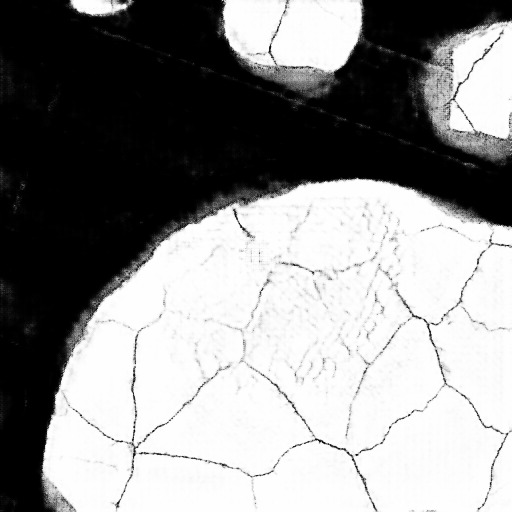}

\end{tabular}

    \end{center}
    \caption{Ablation study. A sample of the dataset demonstrates the importance of each loss term. Compared to the ground truth (GT), the proposed loss (second row) matches very well. Each disabled loss term degrades the prediction quality.}
    \label{fig:ablation_sample}
\end{figure}

The proposed loss consists of several independent loss terms. To showcase that each loss term provides a significant contribution towards the final results, we disable the rendering, parameter, discriminator, and feature loss individually and train the network with the same architecture and training duration described in Section~\ref{subsec:training}. The trained networks are then used to predict every one of the 7175 samples in the test dataset. In Table~\ref{tab:ablation} the mean degradation in quality over the whole test dataset is shown. As seen, every loss term improves the prediction quality. This influence on the quality is displayed in Figure~\ref{fig:ablation_sample}.
Especially the contribution of the adversarial loss $\mathcal{L}_a$ is significant. This is visible in the reconstruction of the fine detail in the metallic map as well as in the lacking detail in the normal map. The parameter and rendering loss help to reduce specular artifacts which are visible in the especially visible in the metallic map. Overall the loss term with the highest influence on prediction quality is the adversarial loss, which suggests that the introduction of an error metric which is not defined on a per-pixel basis is essential in the BRDF estimation quality. This is counter intuitive compared to previous tasks like super-resolution (\cf \cite{enhancenet, Johnson2016}), where the perceptual based losses achieve higher perceptual quality but worse scores \wrt to MSE or PSNR. However, this task is different from super-resolution and can be considered a form of style transfer. The large receptive area of the adversarial loss reduces the chance to run into local minima compared to only leveraging per-pixel information.


\subsection{Network Architecture}
\label{subsec:network_architecture}

The generator architecture is based on \citet{Johnson2016}. However, we use pre-activation residual blocks~\cite{He2016}. The discriminator architecture follows the PatchGAN discriminator from \citet{Isola2017}. Here, both discriminators $D_1$ and $D_2$ (see Figure~\ref{fig:overview}) use the same architecture, but different input resolution. Each discriminator is producing a differently sized output. Each output denotes whether the patch is believed to be from a real or fake sample. During training, for each $512\times512$px input, the $D_1$ output is a $32\times32$ map and for $D_2$ a $16\times16$ map. Hence, each entry from the prediction map has a receptive field of $16\times 16$px or $32\times 32$px area regarding the input image, respectively.

The detailed network architecture is described in the naming convention used by \citet{Johnson2016}.

\paragraph{Generator}
\texttt{c7s1-k} 7x7 Convolution-ReLU with \texttt{k} filters and stride of 1. \texttt{dk} denotes 3x3 Convolution-InstanceNorm-ReLU layer with \texttt{k} filters and stride of 2. \texttt{Rk} denotes a pre-activation residual block with \texttt{k} filters. \texttt{uk} denotes a 3x3 Transposed Convolution with \texttt{k} filters and InstanceNorm with the ReLU activation is used. To further reduce artifacts at the borders a reflection padding is used in every layer of the generator network, except the downscaling \texttt{d} and upscaling layers \texttt{u}:

\texttt{c7s1-64, d128, d256, d512, R512, R512, R512, R512, R512, R512, R512, R512, R512, u256, u128, u64, c7s1-8}.

\paragraph{Discriminator}
\texttt{cn-k} denotes a 4x4 Convolution-LeakyReLU with a stride of 2 and \texttt{k} filters. \texttt{ck} denotes a 4x4 Convolution-InstanceNorm-LeakyReLU with a stride of 2 and \texttt{k} filters. \texttt{cns1-k} denotes a 4x4 Convolution-LeakyReLU with a stride of 1 and \texttt{k} filters. The Leaky\-ReLU uses a slope of 0.2. Both discriminators use the identical architecture:

\texttt{cn-64, c128, c256, c512, cns1-1}.

\subsection{Training}
\label{subsec:training}

The network is trained for 1.000.000 iterations with a batch size of 8 on four Nvidia 1080 Ti GPUs. The network is trained for 200 epochs with 5000 steps per epoch. The Adam optimizer \cite{Kingma2014AdamAM} is used with $\beta_1=0.5$ and $\beta_2=0.999$. 
For the first 100 epochs, the learning rate is set to $\expnumber{2}{-4}$ and afterward linearly decreased to $0$. 
The network architecture, including the rendering loss, is implemented in TensorFlow.

\section{Dataset}
\label{sec:dataset}

\begin{figure}[hbt!]
    \begin{center}
\begin{tabular}{c @{\hskip 0.05in} c @{\hskip 0.2in} c @{\hskip 0.05in} c}
\includegraphics[width=0.22\linewidth]{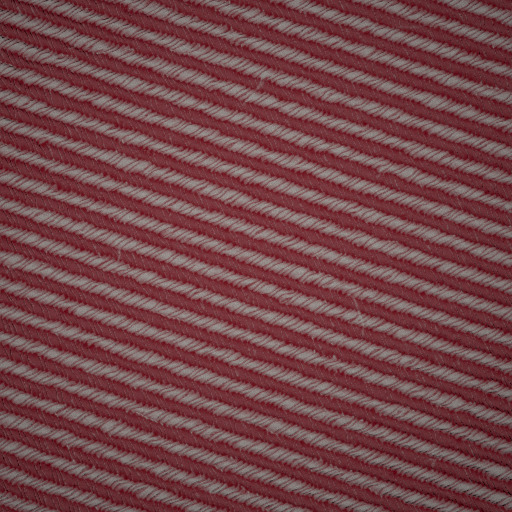} & 
\includegraphics[width=0.22\linewidth]{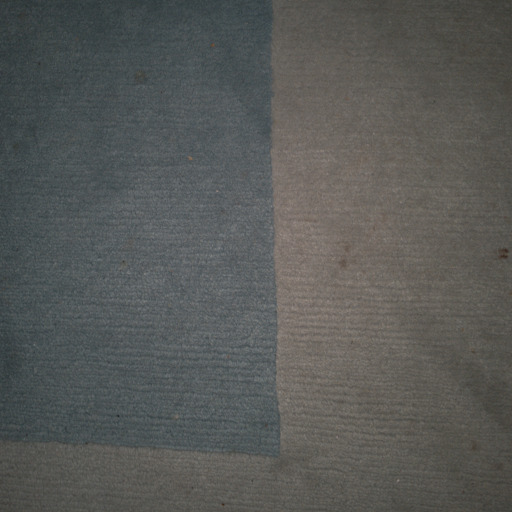} & 
\includegraphics[width=0.22\linewidth]{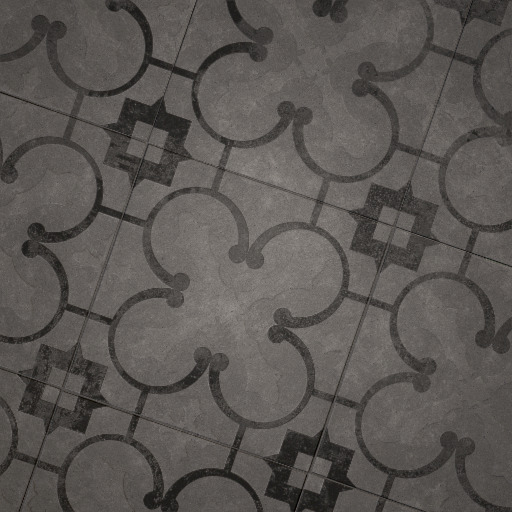} &
\includegraphics[width=0.22\linewidth]{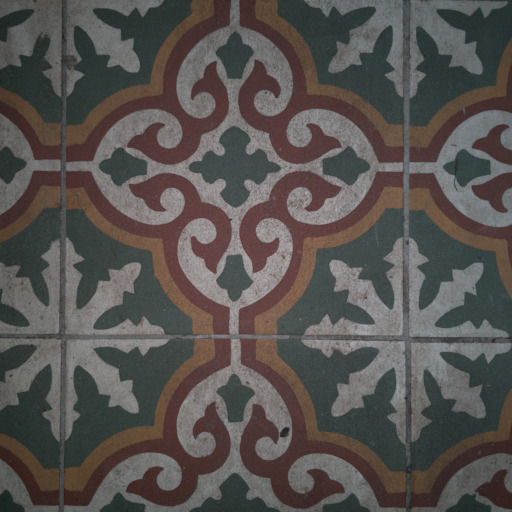} \\

\includegraphics[width=0.22\linewidth]{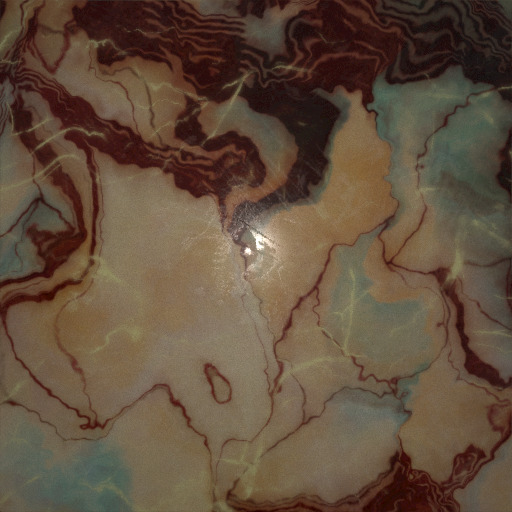} & 
\includegraphics[width=0.22\linewidth]{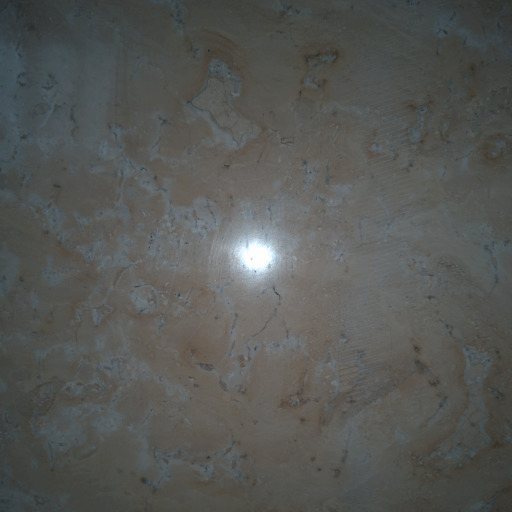} & 
\includegraphics[width=0.22\linewidth]{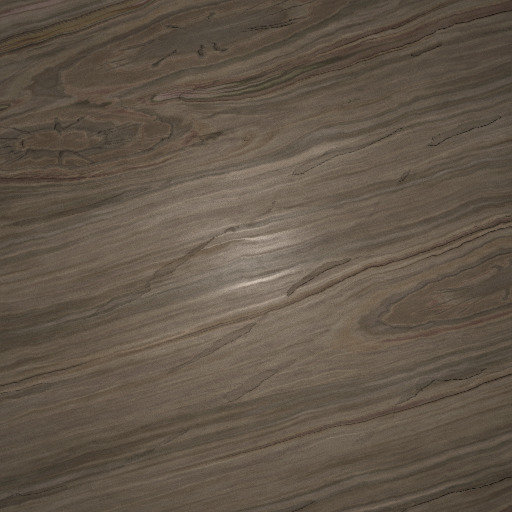} &
\includegraphics[width=0.22\linewidth]{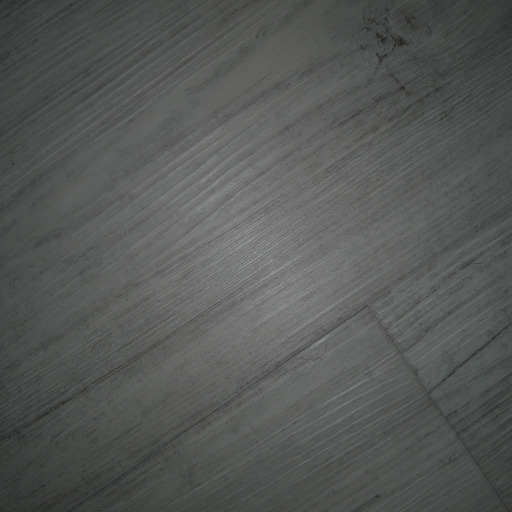}

\end{tabular}

    \end{center}
    \caption{Realistic Dataset. The pairs of similar materials of the synthetic dataset (left) and the real-world (right) indicates that the synthetic set contains realistically looking materials. Samples are visualized in sRGB.}
    \label{fig:comparison_real_dataset}
\end{figure}

\begin{figure*}[hbt!]
    \begin{center}
\begin{tabular}{c @{\hskip 0.0in} c @{\hskip 0.25in} c @{\hskip 0.0in} c @{\hskip 0.25in} c @{\hskip 0.0in} c}
    \includegraphics[width=0.15\linewidth]{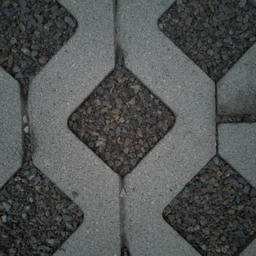} &
    \includegraphics[width=0.15\linewidth]{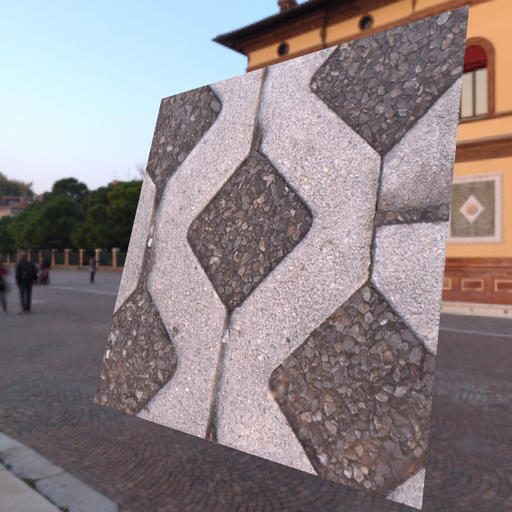} & 
    
    \includegraphics[width=0.15\linewidth]{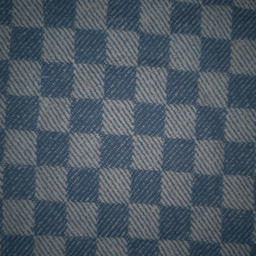} &
    \includegraphics[width=0.15\linewidth]{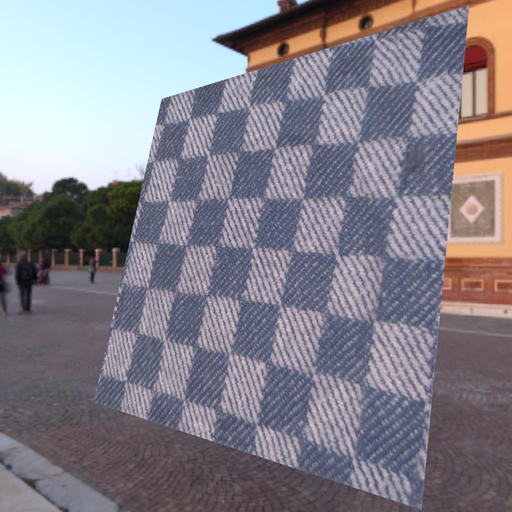} &
    
    \includegraphics[width=0.15\linewidth]{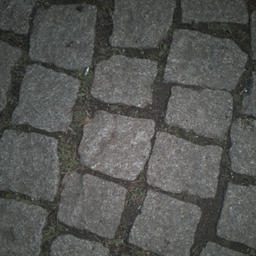} &
    \includegraphics[width=0.15\linewidth]{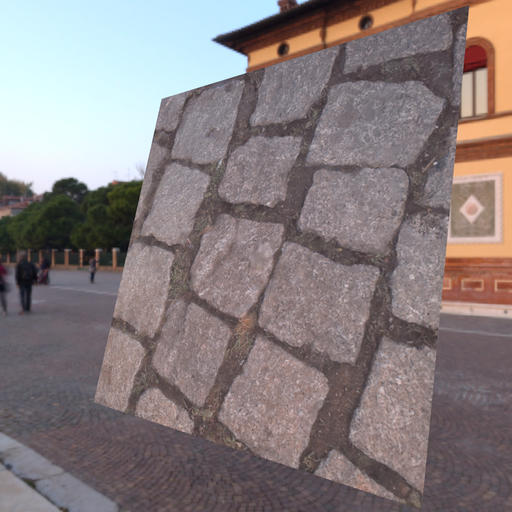}\\
    
    \includegraphics[width=0.15\linewidth]{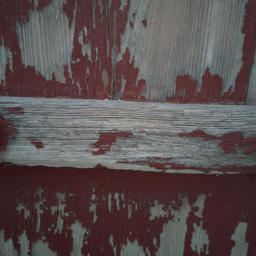} &
    \includegraphics[width=0.15\linewidth]{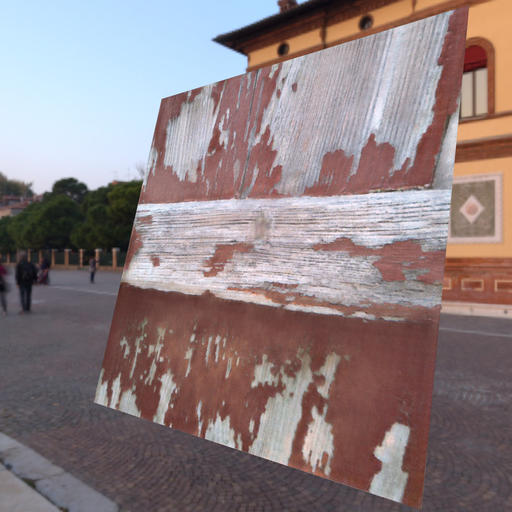} & 
    
    \includegraphics[width=0.15\linewidth]{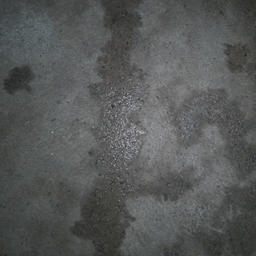} &
    \includegraphics[width=0.15\linewidth]{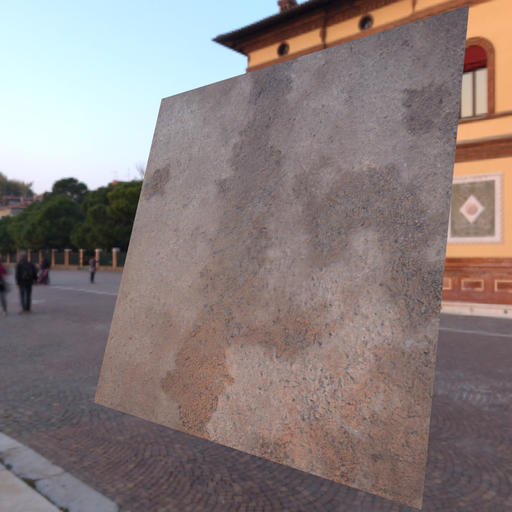} &
    
    \includegraphics[width=0.15\linewidth]{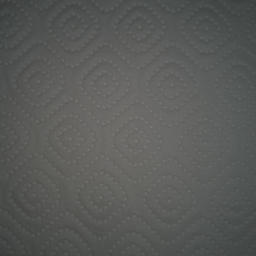} &
    \includegraphics[width=0.15\linewidth]{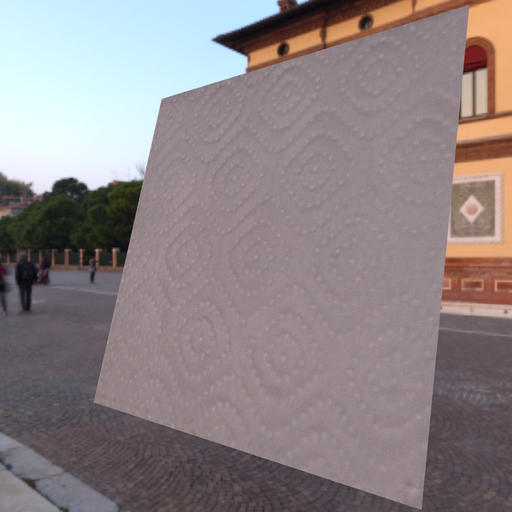}

\end{tabular}

    \end{center}
    \caption{Several real-world photographs with their re-rendered predictions. In each pair the left side is the input flash photograph and on the right is the re-rendered prediction illuminated by an environment map in sunset.}
    \label{fig:real_world_examples}
\end{figure*}

\begin{figure*}[hbt!]
\begin{center}
\begin{tabular}{c @{\hskip 0.2in} c @{\hskip 0.05in} c @{\hskip 0.05in} c @{\hskip 0.05in} c @{\hskip 0.2in} c}
     \includegraphics[width=0.15\linewidth]{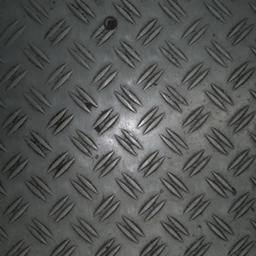} & 
     \includegraphics[width=0.15\linewidth]{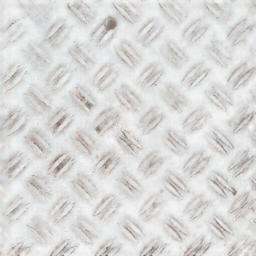} & 
     \includegraphics[width=0.15\linewidth]{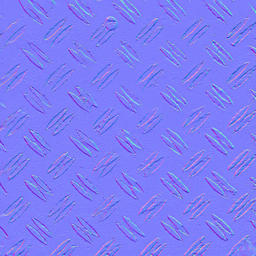} &
     \includegraphics[width=0.15\linewidth]{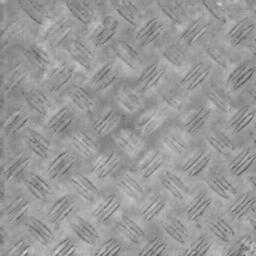} & 
     \includegraphics[width=0.15\linewidth]{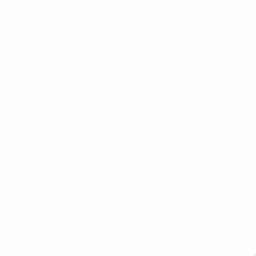} & 
     \includegraphics[width=0.15\linewidth]{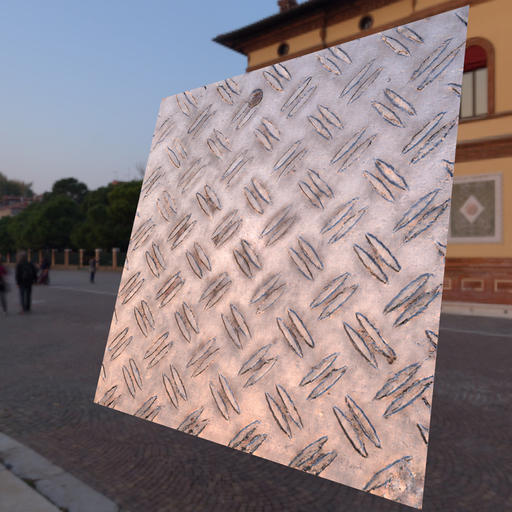} \\
     
     \includegraphics[width=0.15\linewidth]{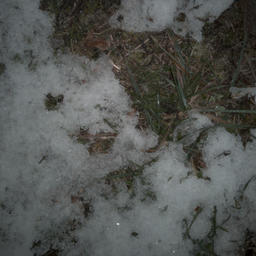} & 
     \includegraphics[width=0.15\linewidth]{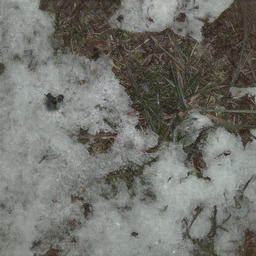} & 
     \includegraphics[width=0.15\linewidth]{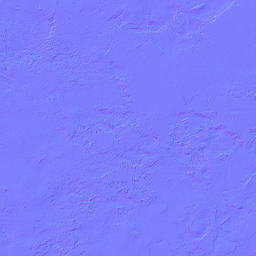} &
     \includegraphics[width=0.15\linewidth]{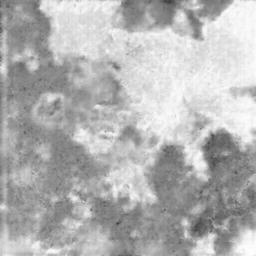} & 
     \includegraphics[width=0.15\linewidth]{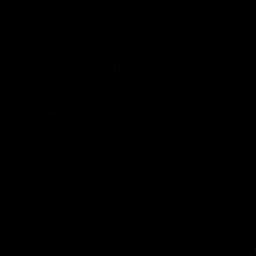} & 
     \includegraphics[width=0.15\linewidth]{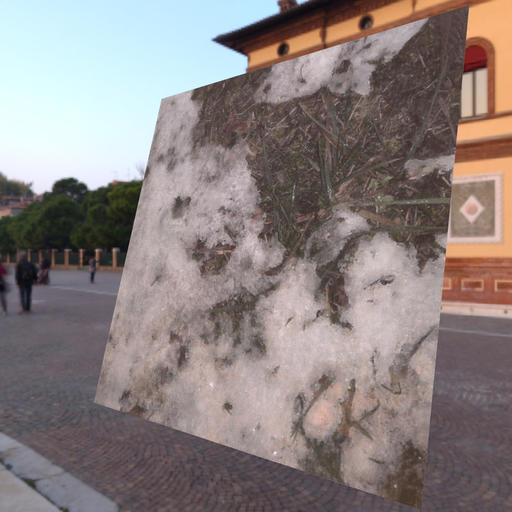} \\
     
     \includegraphics[width=0.15\linewidth]{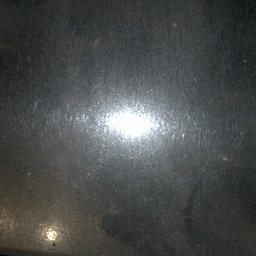} & 
     \includegraphics[width=0.15\linewidth]{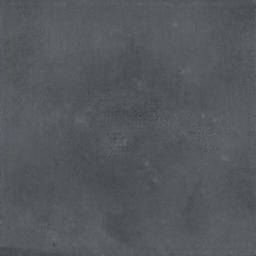} & 
     \includegraphics[width=0.15\linewidth]{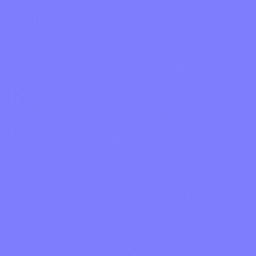} &
     \includegraphics[width=0.15\linewidth]{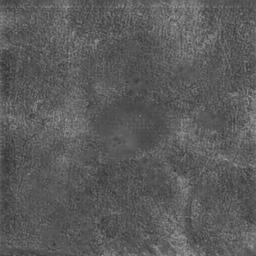} & 
     \includegraphics[width=0.15\linewidth]{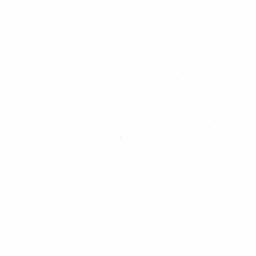} & 
     \includegraphics[width=0.15\linewidth]{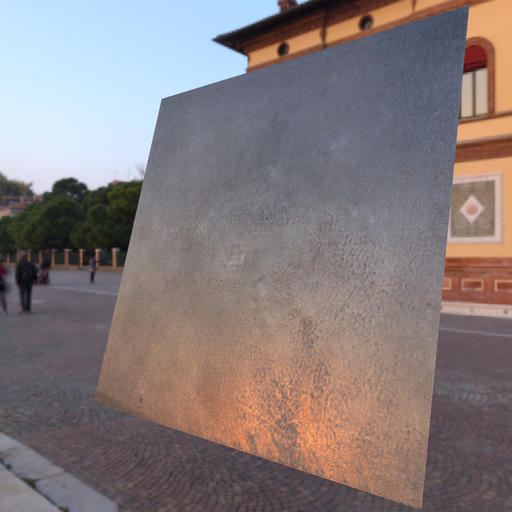} \\
     
     Input & Base color & Normal & Roughness & Metallic & Re-rendered
\end{tabular}

\end{center}
    \caption{Recovered SVBRDF parameters of specular materials from real-world flash photographs (left). The specular behavior of each material is captured well. Especially the middle example showcases the high degree of detail in the estimation. Here, the difference in the roughness between the slightly frozen snow and the grass and dirt is resolved well. The bottom example demonstrates that additional light sources from the environment are successfully removed.}
    \label{fig:real_world_specular_example}
\end{figure*}

The dataset is generated with Allegorithmic Substance software and the Substance Share dataset~\cite{allegorithmic}. Community members can upload own procedural materials and rate the uploaded materials. By leveraging the rating system, 175 of the highest rated materials are gathered. From these materials, the list of adjustable parameters is collected. Depending on the number of adjustable parameters up to 100 permutations are generated per material. Each output is generated with all parameters being randomly drawn from a normal distribution with $\mu$ being the default value, and $\sigma$ is set to sample the whole specified value range for the parameter. This ensures that the materials are generated with plausible parameters. 
The materials are then exported to 2048x2048 pixel parameter maps. These are randomly cropped, rotated and scaled to $512\times512$ pixel resolution seven times. An 80:20 split
into training and test data is applied on the Substance files before permutation and post-processing. Overall, this leaves 40544 samples for training and 7175 samples for test.

In Figure~\ref{fig:comparison_real_dataset} several real-world and dataset samples are compared. As seen, the rendered samples from the dataset are hard to distinguish from the real-world ones whenn applying the following processing steps, demonstrating the quality of the synthetic data set:

For the input, each material is rendered three times with a randomly rotated High Dynamic Range (HDR) environment map from a pool of 20 outdoor and indoor maps taken from HDRI Haven~\cite{hdrihaven}. The test dataset is only rendered once from a pool of 6 environment maps. The rendering is done in Mitsuba~\cite{mitsuba} with 196 samples per pixel. The HDR output of the Mitsuba renderer is converted to Low Dynamic Range (LDR) with an auto exposure algorithm. This algorithm calculates an exposure scalar for the photograph. The exact algorithm is outlined in the Appendix~\ref{sec:append_auto_exposure}. No additional tone mapping is applied, and the images are left in linear color space. This step is taken to match raw mobile phone images exported to a linear color space.

\section{Evaluation}

The evaluation is performed both on real-world photographs demonstrating the capability to operate on non-synthetic data, and on synthetic data with known ground truth to quantify the error of the estimation. To match real-world conditions the test set is rendered with a randomly rotated environment map, varying flash strengths and color temperatures. It is then passed to the auto exposure algorithm (see appendix) and thus converted to LDR. The test dataset is used in this setup, and thus we can compare our proposed method against \shortcite{Deschaintre2018} and \shortcite{Li2017}.

\subsection{Real-world Photographs}

Two mobile phones, a Google Pixel 2 and a Samsung Galaxy S9, are used to capture real-world data. The images are captured in RAW to minimize post-processing by the mobile phone. The mobile phone is held parallel to the surface with a distance of approximately 50cm, and a camera flash is required for the capture. With this setup, 282 casual photographs of surfaces are captured. The RAW images are then exported in linear color space and cropped with a 1:1 aspect ratio with the camera flash being roughly at the center. Our network is capable of extracting high-quality parameter maps for various surfaces even with multiple mixed materials. In Figure~\ref{fig:teaser} various estimated materials are rendered on a complex model. The corresponding input image is shown in the upper left corner. As seen the reflectance behavior is estimated realistically and matches the input material well. In Figure~\ref{fig:real_world_specular_example} several challenging surfaces are highlighted. Especially the detail in the reconstructed parameters are noteworthy. For example in the second row, the fine surface normal and base color detail for the snow and the different roughness for the snow and grass patches is difficult to predict. The first row highlights that the harsh specular reflection of the camera flash is removed successfully and no artifacts are visible in the parameters. The prediction results in a believable surface for the metal plate.

The last row highlights the importance of training with environment lit renderings. In the input image, an additional light source is visible in the bottom left. However, in the reconstruction, this highlight is removed fully, because the network learned that these light sources are not a reasonable feature of the material. Various other materials are evaluated in Figure~\ref{fig:real_world_examples}. Here, the input image is on the left of each pair and the rendered predictions on the right. Every material realistically reproduces the behavior of each surface. Additional materials and a video showing comparisons under novel viewing and lighting conditions are available in the supplementary.

\subsection{Comparison}

\begin{figure}[hbt!]
    \centering
    \includegraphics[width=0.8\linewidth]{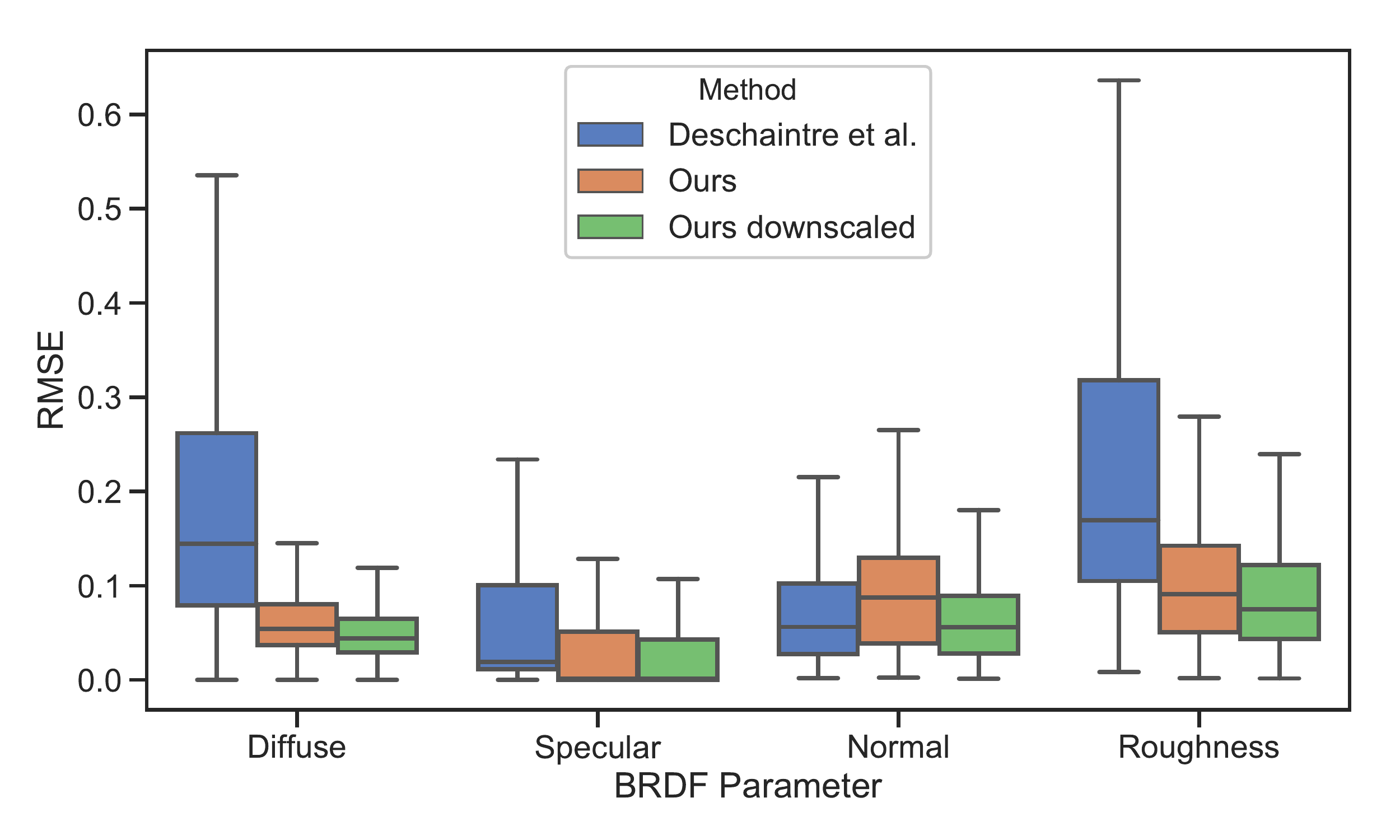}
    \caption{Comparison to the work of \citet{Deschaintre2018} on our 7175 test materials. As our approach predicts at higher resolution, we compare our results against $512\times512$px resolution ground truth images and additionally downscale our results with the ground truth to $256\times256$ px resolution to match the resolution used in \citet{Deschaintre2018}. This way parameters with a high degree of fine detail like the normal map can be compared similarly. Our approach is trained to cope with natural illumination besides the flash light, and this is also true for the test set. Therefore on can note the significantly more precise estimation of the diffuse and the roughness channel in our framework. }
    \label{fig:rmse_comparison}
\end{figure}

\begin{figure*}[hbt!]
\begin{tabular}{r @{\hskip 0.05in} c @{\hskip 0.05in} c @{\hskip 0.05in} c @{\hskip 0.05in} c @{\hskip 0.05in} c @{\hskip 0.05in} c}

    \raisebox{2.7\normalbaselineskip}[0pt][0pt]{\rotatebox[origin=c]{90}{\parbox{0pt}{GT}}} & 
    &
    \includegraphics[width=0.15\textwidth]{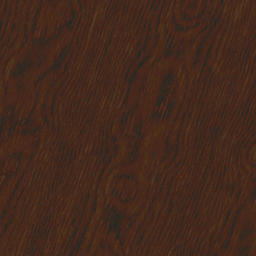} &
    \includegraphics[width=0.15\textwidth]{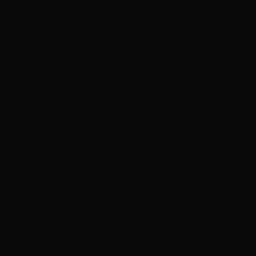} &
    \includegraphics[width=0.15\textwidth]{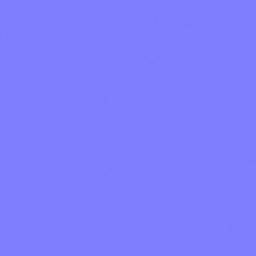} &
    \includegraphics[width=0.15\textwidth]{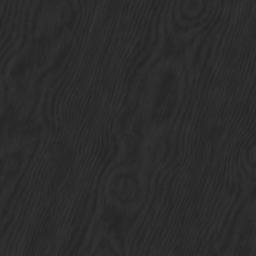} &
    \includegraphics[width=0.15\textwidth]{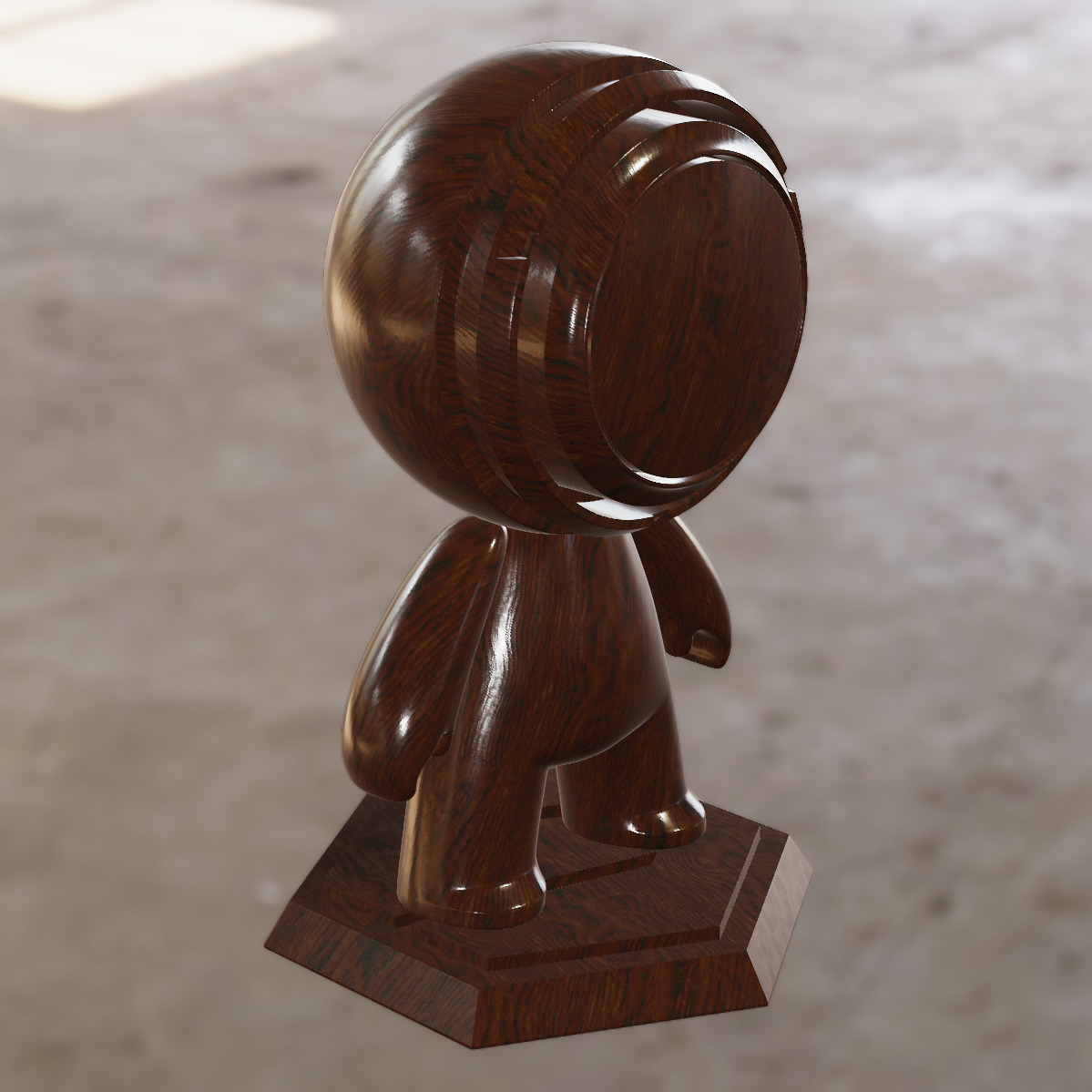} \\

    \raisebox{0.75\normalbaselineskip}[0pt][0pt]{\rotatebox[origin=c]{90}{\parbox{0pt}{\mbox{[Li et al. 2017]} }}} & 
    \includegraphics[width=0.15\textwidth]{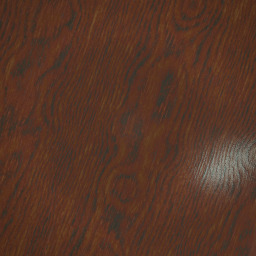} &
    \includegraphics[width=0.15\textwidth]{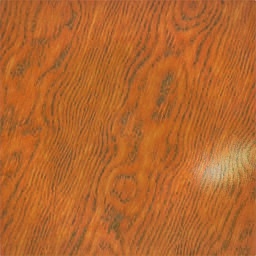} &
    \includegraphics[width=0.15\textwidth]{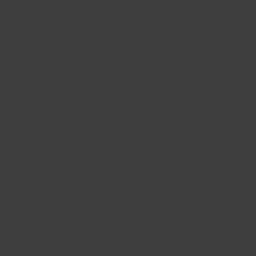} &
    \includegraphics[width=0.15\textwidth]{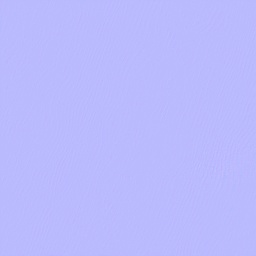} &
    \includegraphics[width=0.15\textwidth]{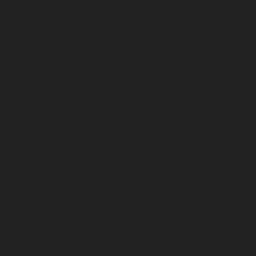} &
    \includegraphics[width=0.15\textwidth]{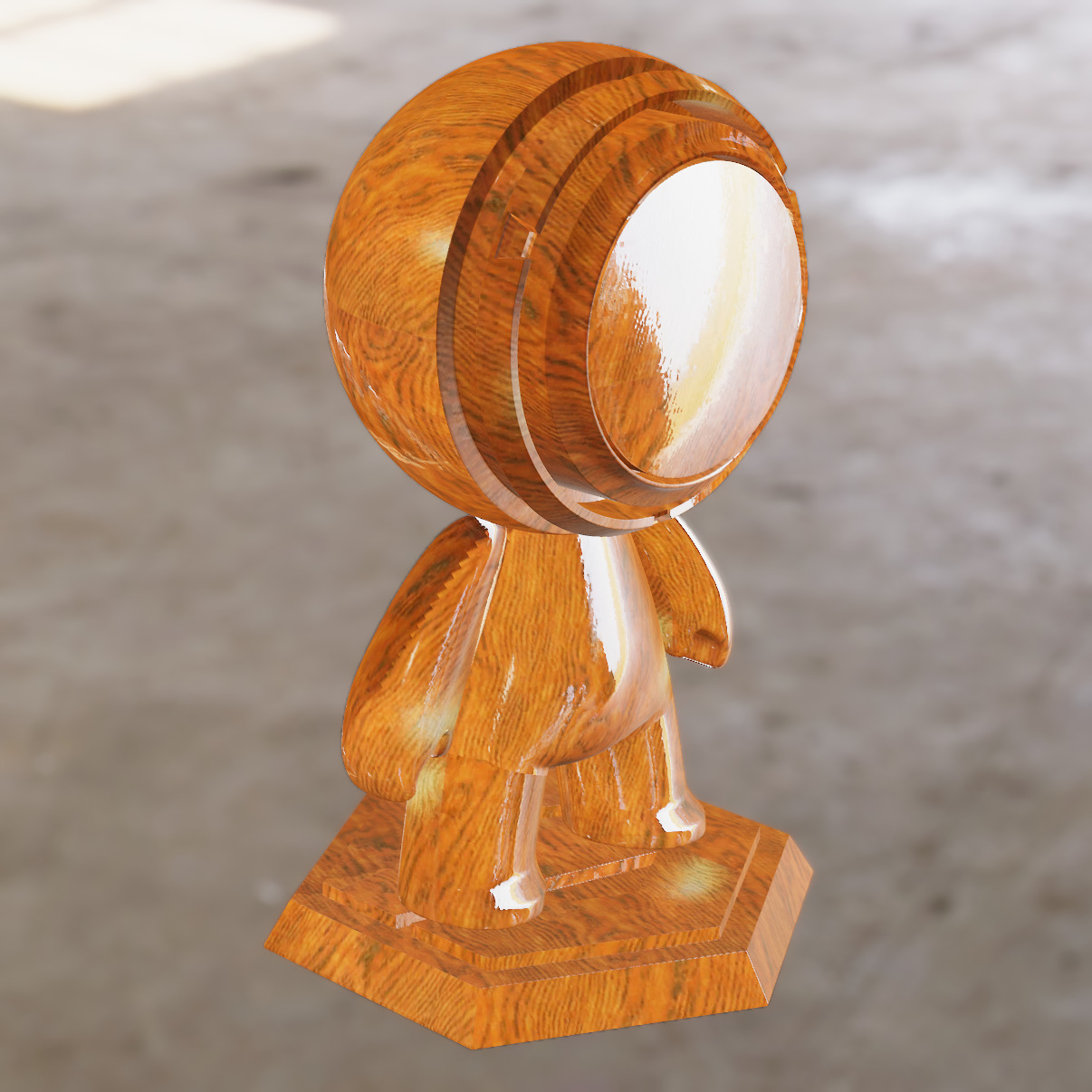} \\

    \raisebox{0\normalbaselineskip}[0pt][0pt]{\rotatebox[origin=c]{90}{\parbox{0pt}{\mbox{[Deschaintre et al.]} }}} & 
    \includegraphics[width=0.15\textwidth]{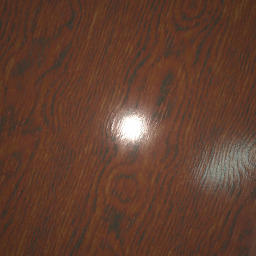} &
    \includegraphics[width=0.15\textwidth]{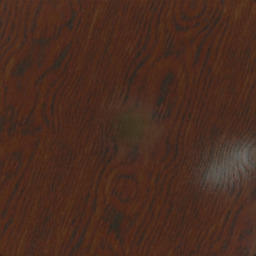} &
    \includegraphics[width=0.15\textwidth]{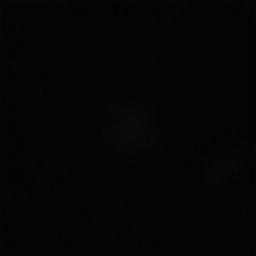} &
    \includegraphics[width=0.15\textwidth]{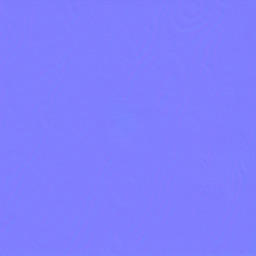} &
    \includegraphics[width=0.15\textwidth]{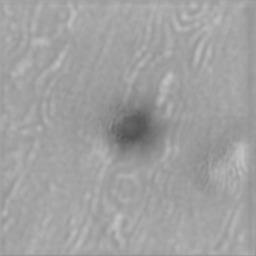} &
    \includegraphics[width=0.15\textwidth]{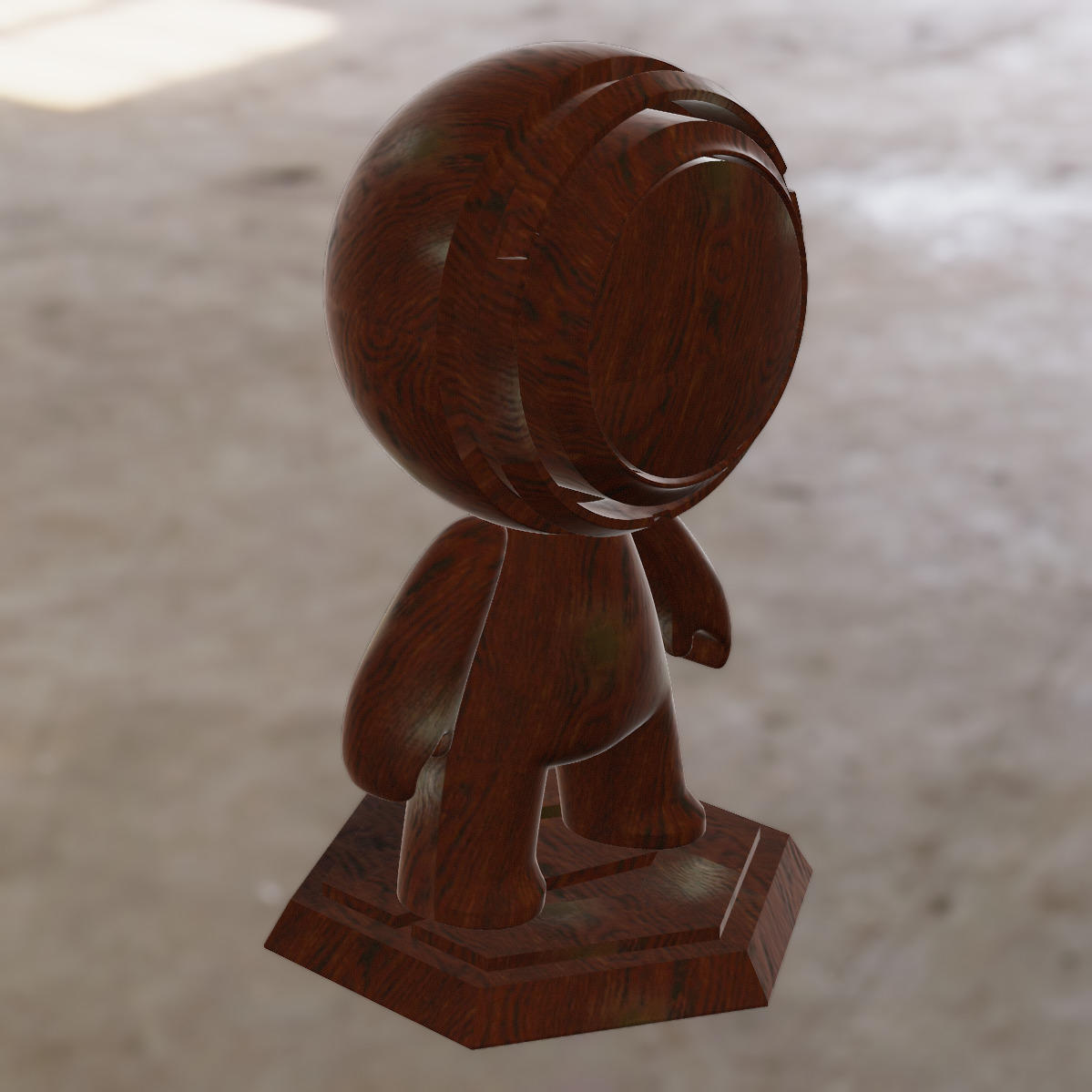} \\

    \raisebox{3\normalbaselineskip}[0pt][0pt]{\rotatebox[origin=c]{90}{Ours}} & 
    \includegraphics[width=0.15\textwidth]{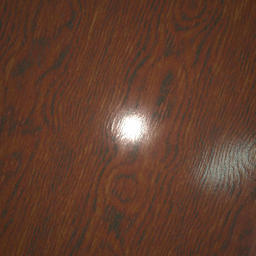} &
    \includegraphics[width=0.15\textwidth]{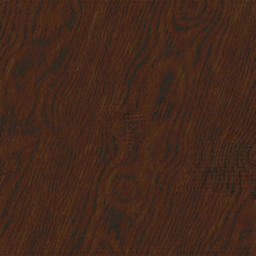} &
    \includegraphics[width=0.15\textwidth]{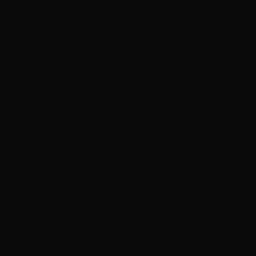} &
    \includegraphics[width=0.15\textwidth]{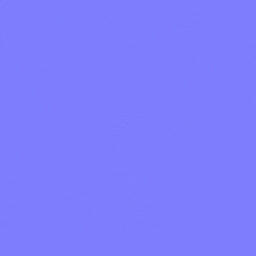} &
    \includegraphics[width=0.15\textwidth]{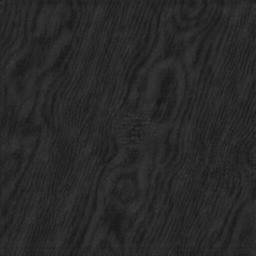} &
    \includegraphics[width=0.15\textwidth]{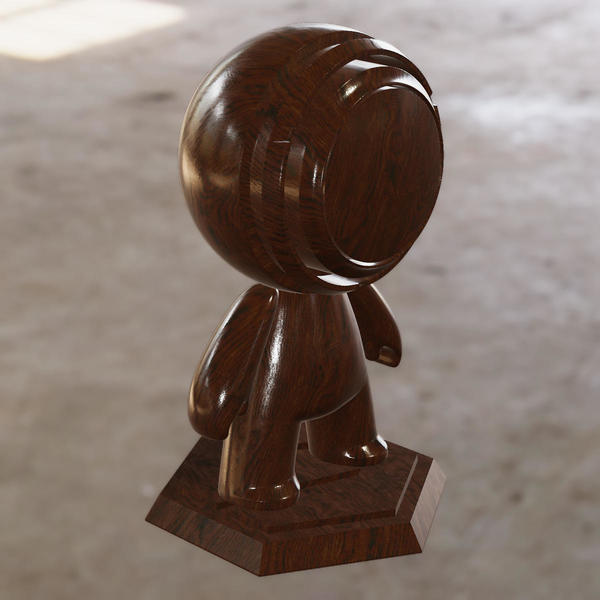} \\

    & Input & Diffuse & Specular & Normal & Roughness & Rendered
\end{tabular}

\Description{Visual comparison between the results on synthetic data between Deschaintre \etal, Li \etal [2017] and our approach.} 
  \caption{Comparison between \citet{Deschaintre2018}, \citet{Li2017} and our approach on synthetic data. The input image with the ground truth is located in the top row. Followed by the prediction of Li \citet{Li2017}, Deschaintre \citet{Deschaintre2018} and ours. Li and Deschaintre predict on $256\x256$px resolution, while our approach processes $512\x512$px resolution. It is worth noting that \citet{Li2017} predicts BRDF only using passive illumination and the specular and roughness parameters are only homogenous. Our approach is the only method which reliably removes the secondary highlight due to the environment and correctly estimates the roughness. }
  \label{fig:wood_comparison}
\end{figure*}

\begin{figure*}[hbt!]
  \begin{center}
\begin{tabular}{r @{\hskip 0.05in} c @{\hskip 0.05in} c @{\hskip 0.05in} c @{\hskip 0.05in} c @{\hskip 0.05in} c @{\hskip 0.05in} c}

    \raisebox{2.7\normalbaselineskip}[0pt][0pt]{\rotatebox[origin=c]{90}{\parbox{0pt}{GT}}} & 
    &
    \includegraphics[width=0.15\textwidth]{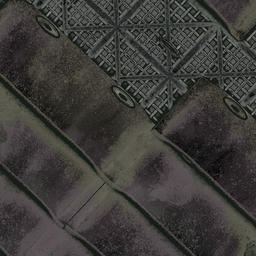} &
    \includegraphics[width=0.15\textwidth]{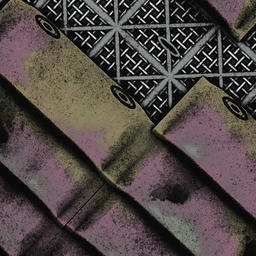} &
    \includegraphics[width=0.15\textwidth]{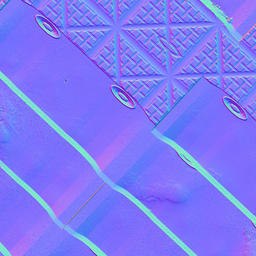} &
    \includegraphics[width=0.15\textwidth]{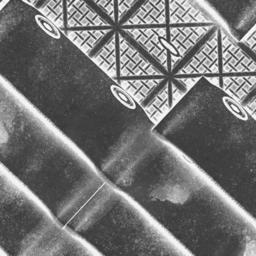} &
    \includegraphics[width=0.15\textwidth]{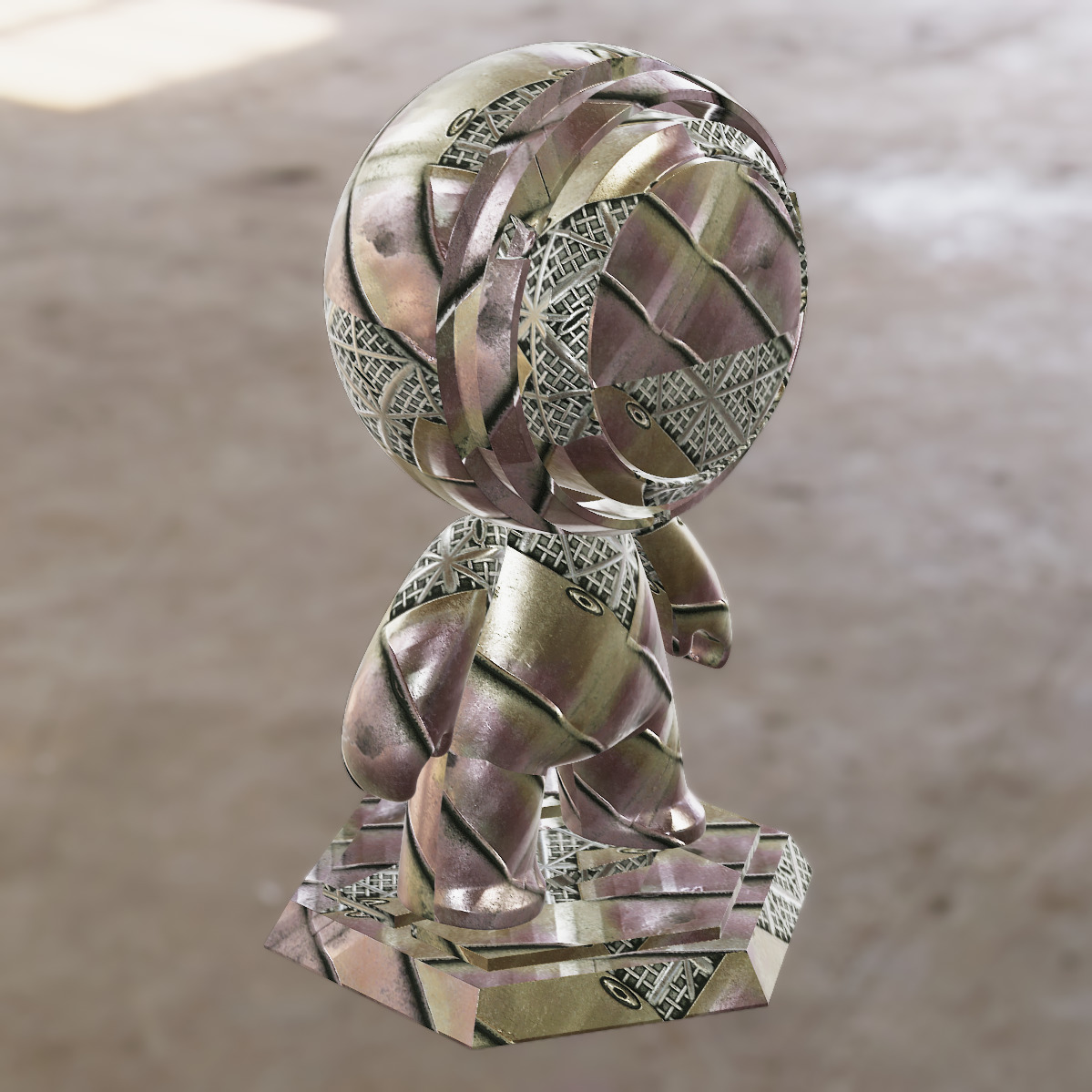} \\

    \raisebox{0.75\normalbaselineskip}[0pt][0pt]{\rotatebox[origin=c]{90}{\parbox{0pt}{\mbox{[Li et al. 2017]} }}} & 
    \includegraphics[width=0.15\textwidth]{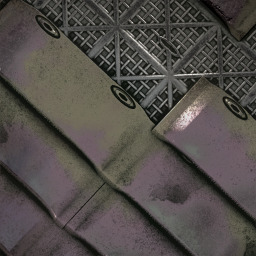} &
    \includegraphics[width=0.15\textwidth]{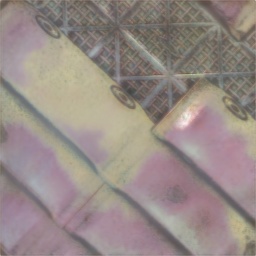} &
    \includegraphics[width=0.15\textwidth]{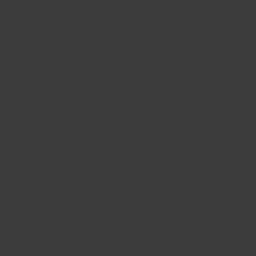} &
    \includegraphics[width=0.15\textwidth]{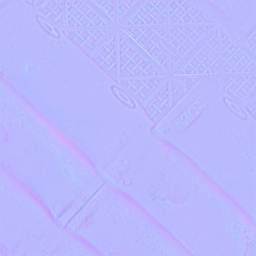} &
    \includegraphics[width=0.15\textwidth]{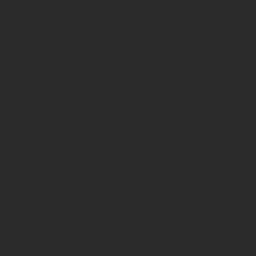} &
    \includegraphics[width=0.15\textwidth]{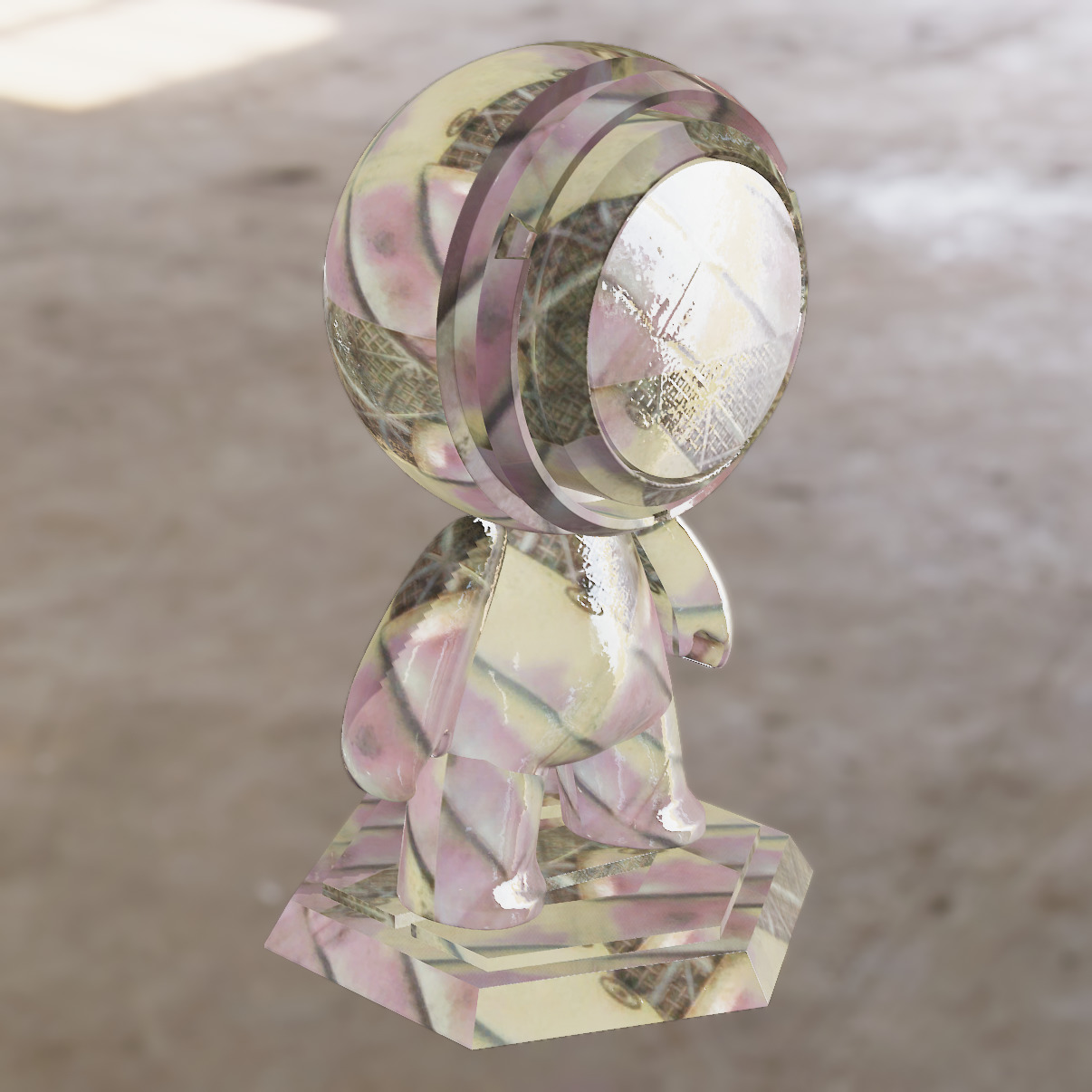} \\

    \raisebox{0\normalbaselineskip}[0pt][0pt]{\rotatebox[origin=c]{90}{\parbox{0pt}{\mbox{[Deschaintre et al.]} }}} & 
    \includegraphics[width=0.15\textwidth]{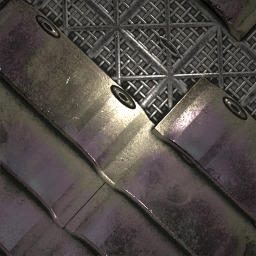} &
    \includegraphics[width=0.15\textwidth]{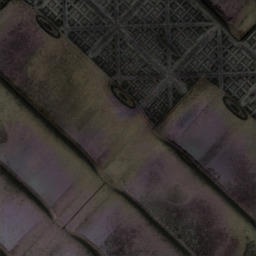} &
    \includegraphics[width=0.15\textwidth]{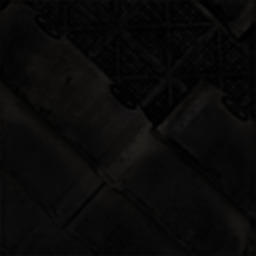} &
    \includegraphics[width=0.15\textwidth]{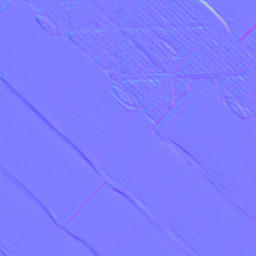} &
    \includegraphics[width=0.15\textwidth]{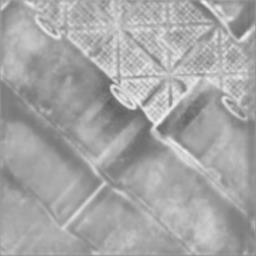} &
    \includegraphics[width=0.15\textwidth]{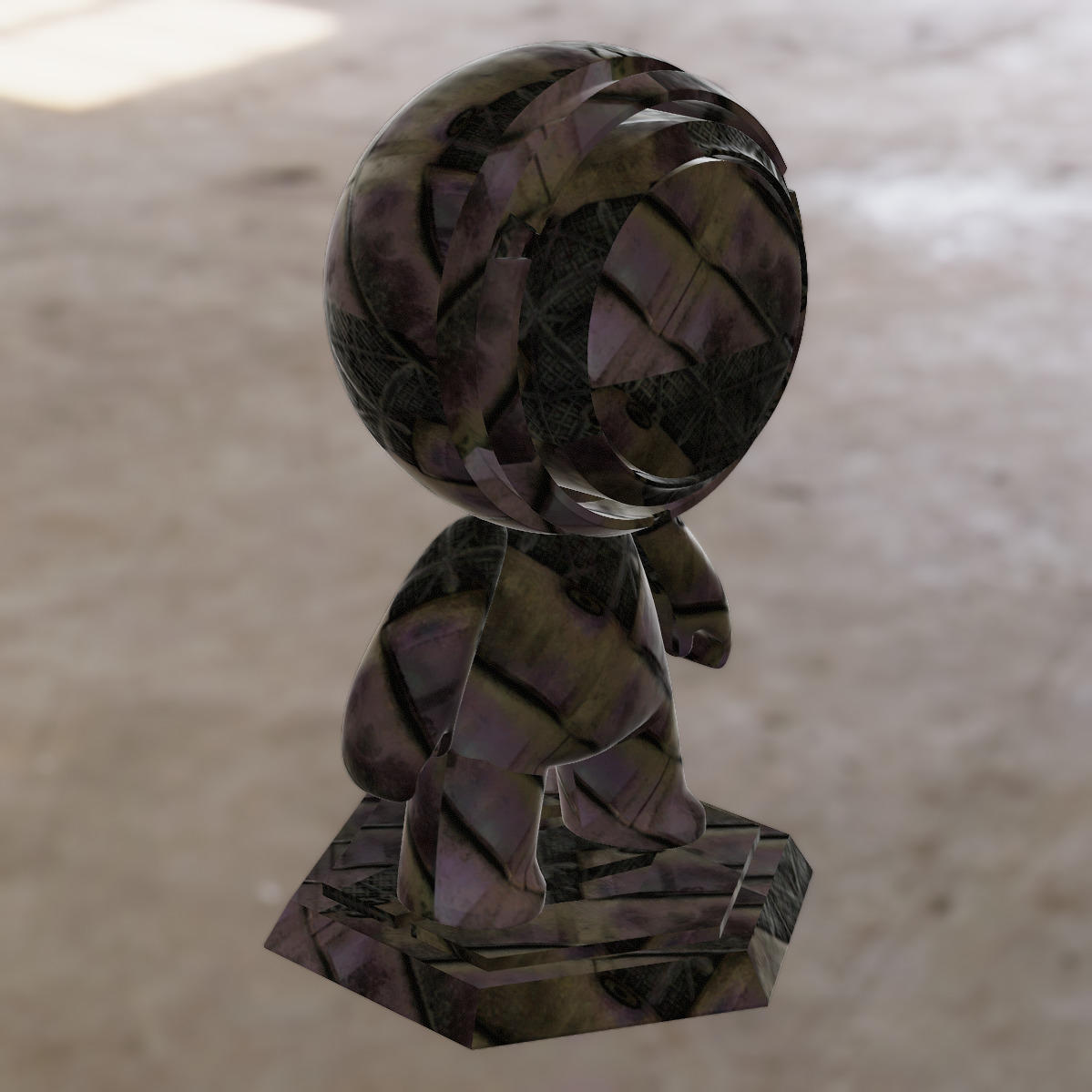} \\

    \raisebox{3\normalbaselineskip}[0pt][0pt]{\rotatebox[origin=c]{90}{Ours}} & 
    \includegraphics[width=0.15\textwidth]{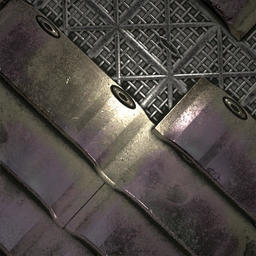} &
    \includegraphics[width=0.15\textwidth]{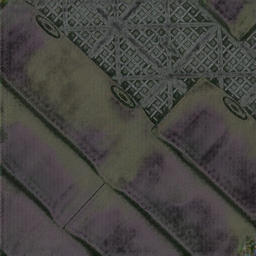} &
    \includegraphics[width=0.15\textwidth]{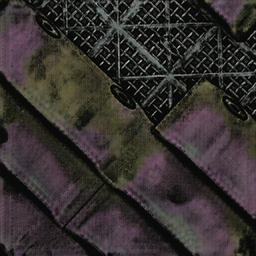} &
    \includegraphics[width=0.15\textwidth]{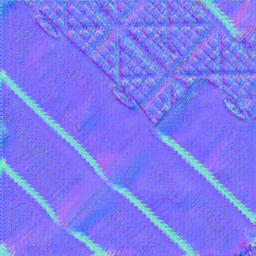} &
    \includegraphics[width=0.15\textwidth]{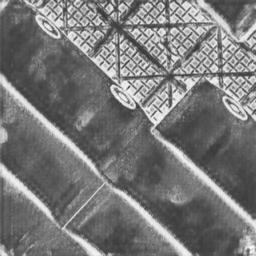} &
    \includegraphics[width=0.15\textwidth]{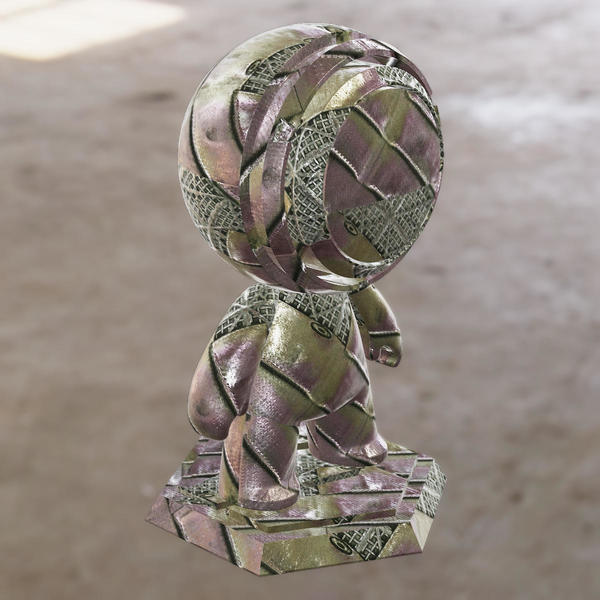} \\

    & Input & Diffuse & Specular & Normal & Roughness & Rendered
\end{tabular}

\Description{Further comparisons between Deschaintre \etal, Li \etal and our approach on a metallic material.}
  \end{center}
  \caption{Further comparison between \citet{Deschaintre2018}, \citet{Li2017} and our approach on synthetic data. In this example our approach is able to predict the specular component including the roughness significantly more precisely. The same notes regarding \citet{Li2017} from Fig. \ref{fig:wood_comparison} apply here.}
  \label{fig:metal_comparison}
\end{figure*}

To correctly quantify the error made by the estimation, the test dataset is used. This dataset, consisting of 7175 materials, is rendered with previously unseen environment maps. Because the ground truth parameter is known in this case, the error can be calculated. Additionally, we can compare our method against the method of \citet{Deschaintre2018}. Deschaintre \etal trained their method on a similar dataset generated from mostly the same Substance materials from Substance Share (\cf \cite{Deschaintre2018}). Additonally, both methods use the Cook-Torrance BRDF model, but the parametrization is slightly different. In section \ref{sec:material_representation} the extraction of the specular and diffuse albedo color from the basecolor and metallic parameters is explained. It is noteworthy that the method of Deschaintre \etal only trained on $256\times256$px resolution images. Less detail needs to be reconstructed by their method. Our method, on the other hand, provides a $512\times512$ pixel resolution output. As \citet{Li2017} try to solve a different problem, the estimation from BRDF parameters from only passive illumination, and additionally only estimate spatially-varying diffuse and surface normal parameters, we only show results from then in a visual comparison (see Fig. \ref{fig:wood_comparison} and \ref{fig:metal_comparison}).

In Figure~\ref{fig:rmse_comparison} we compare both methods on the test dataset. \citet{Deschaintre2018} and our approach receive the input in their respective native resolution. Our method shows better performance in full resolution by a large margin in nearly every parameter except the surface normal. This is due to the fine detail being hardly visible in the rendered image. By scaling our result to $256\times256$px and comparing against ground truth parameters in the same resolution, our method now achieves even better performance in every parameter. This is shown in Figure~\ref{fig:rmse_comparison} as the method 'Ours downscaled'. This experiment only shows that their approach is not robust to environmental illumination, while they achieve much better results on their own test data which only contains the flash light. 

Visually this can be seen in Figure~\ref{fig:wood_comparison} and \ref{fig:metal_comparison}. In Figure~\ref{fig:wood_comparison} a non-metallic material with a harsh secondary illumination is shown. Our method removes the secondary highlight successfully while a strong artifact from the specular highlight is visible in the roughness parameter in the prediction of \citet{Deschaintre2018} and \citet{Li2017}. Our method produces parameters which capture the hand-authored style and are close to the ground truth. In the metallic example seen in Figure~\ref{fig:metal_comparison} our method captures the detail and general color of the material well. 

\begin{figure*}[hbt!]
    \centering
    \begin{tabular}{l c @{\hskip 0.05in} c @{\hskip 0.05in} c @{\hskip 0.05in} c}

    & Galaxy S8 &  One Plus 5 & One Plus 6 & Xiaomi Mi Mix 2 \\

    Deschaintre input& 
    \includegraphics[width=0.15\textwidth]{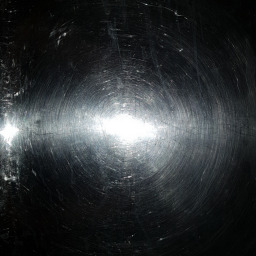} &
    \includegraphics[width=0.15\textwidth]{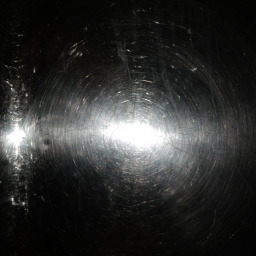} &
    \includegraphics[width=0.15\textwidth]{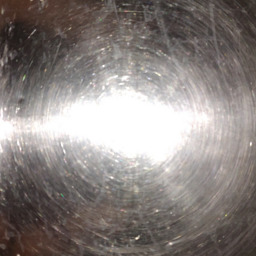} &
    \includegraphics[width=0.15\textwidth]{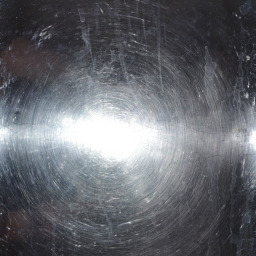} \\

   Deschaintre Rerender & 
    \includegraphics[width=0.15\textwidth]{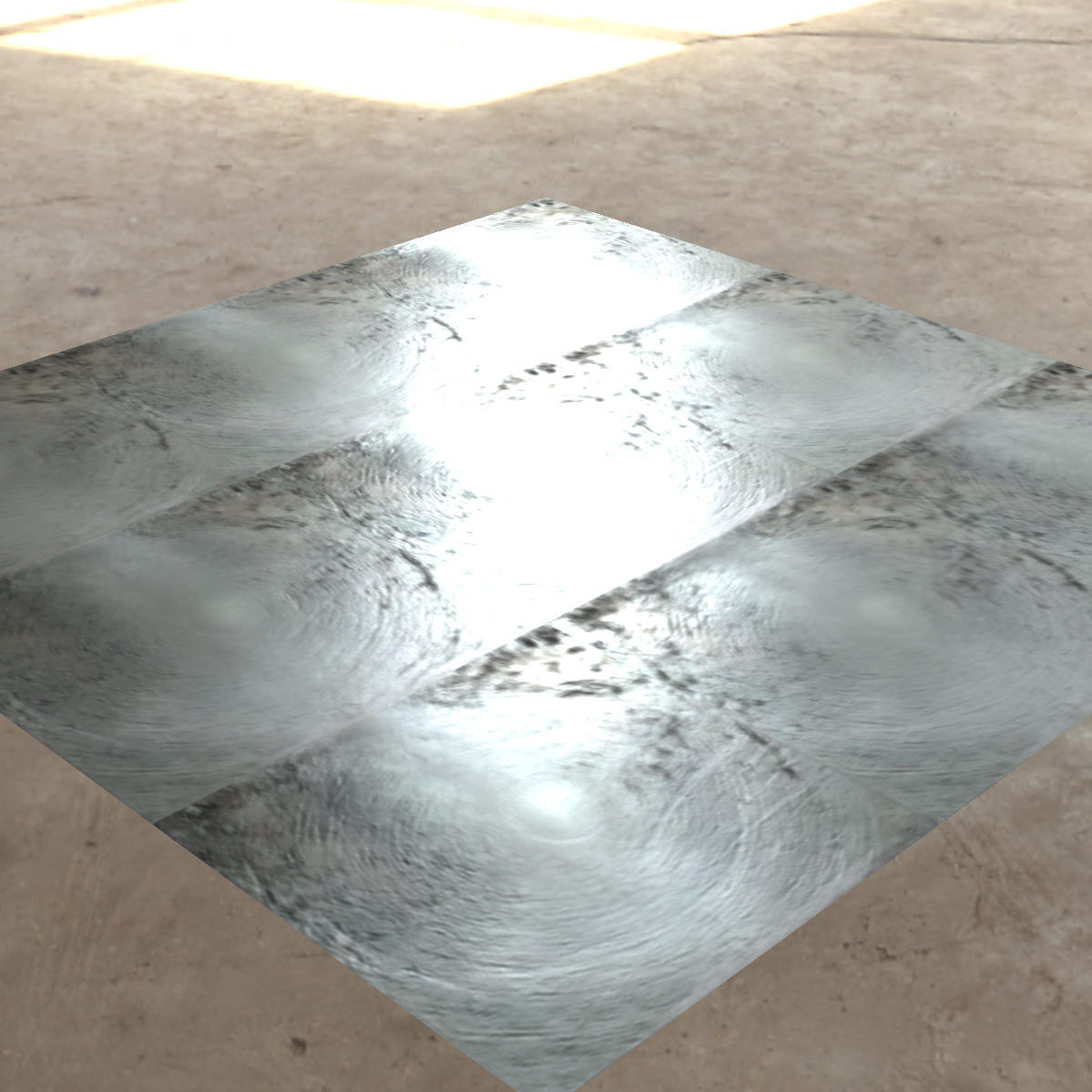} &
    \includegraphics[width=0.15\textwidth]{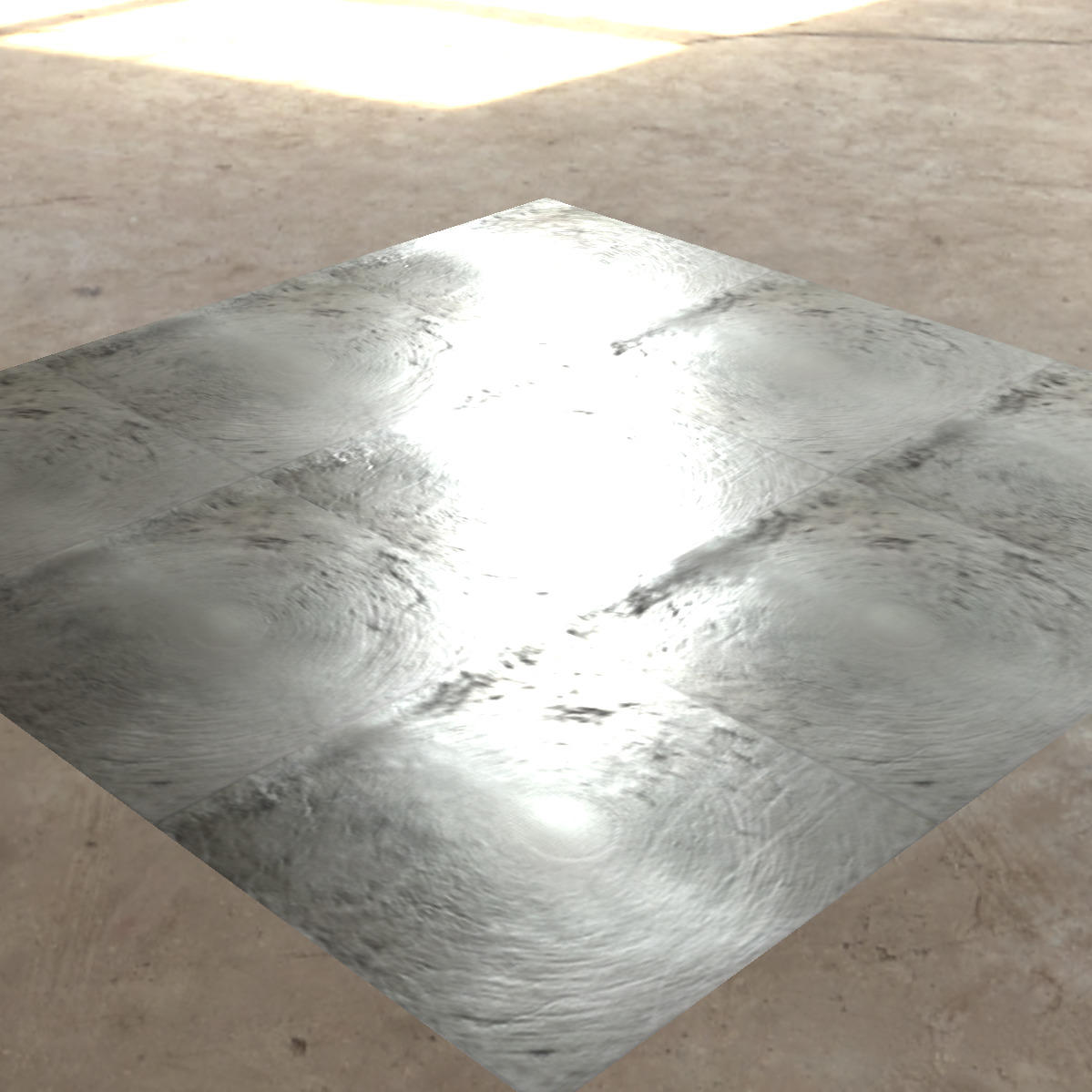} &
    \includegraphics[width=0.15\textwidth]{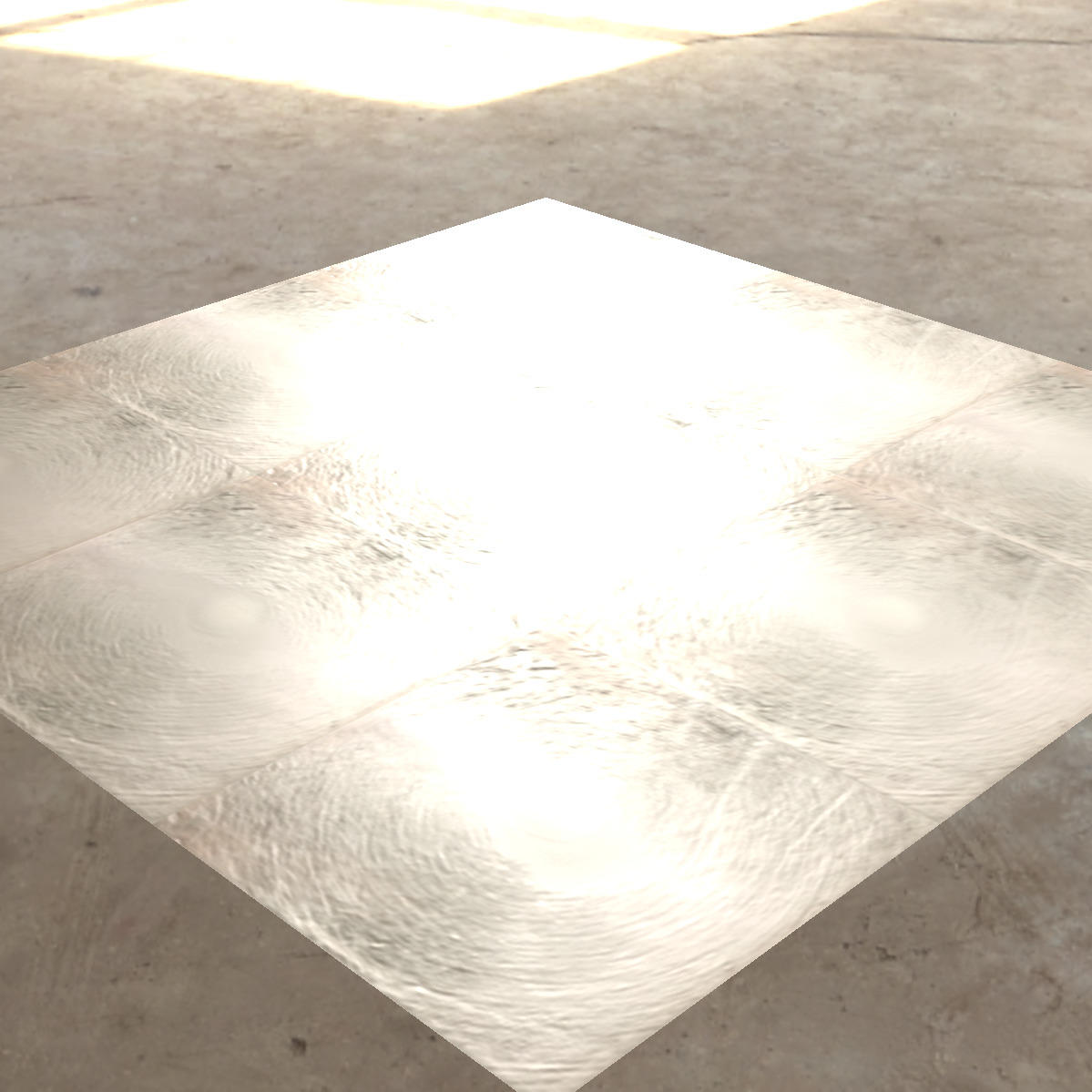} &
    \includegraphics[width=0.15\textwidth]{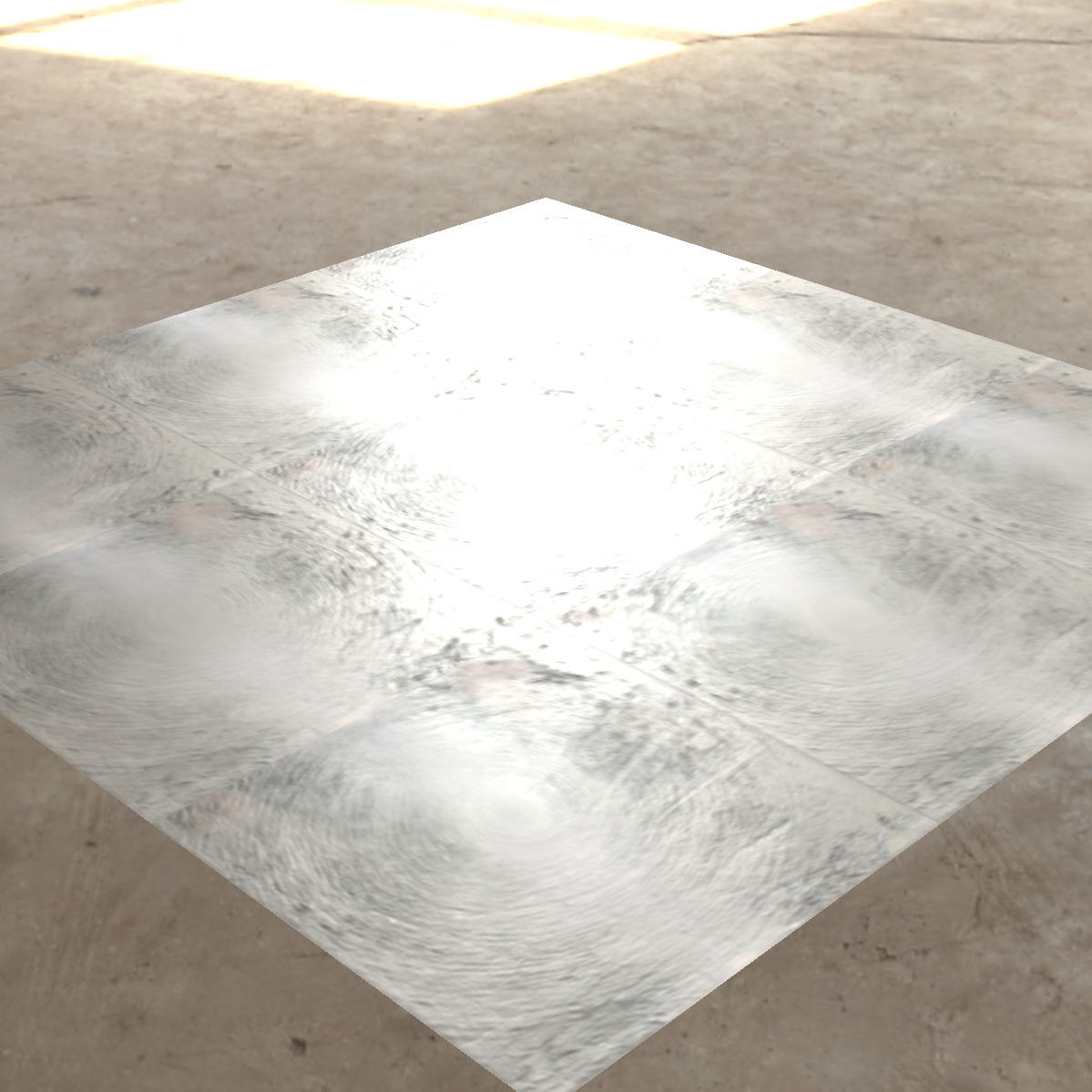} \\

    Ours input & 
    \includegraphics[width=0.15\textwidth]{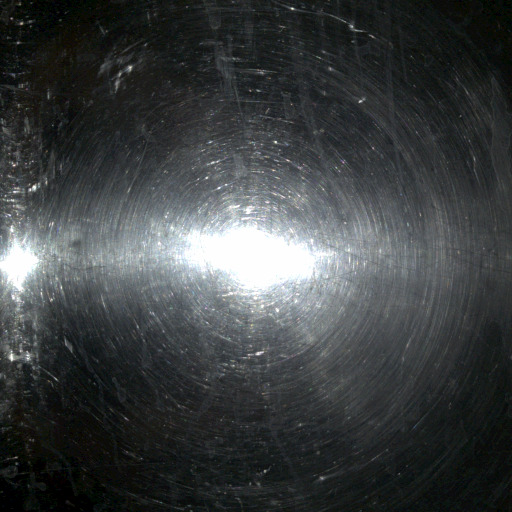} &
    \includegraphics[width=0.15\textwidth]{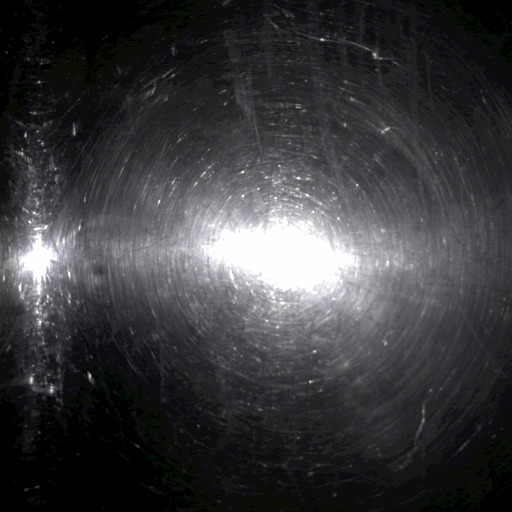} &
    \includegraphics[width=0.15\textwidth]{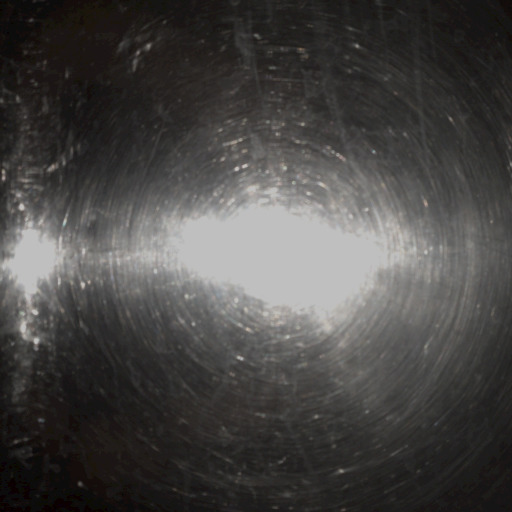} &
    \includegraphics[width=0.15\textwidth]{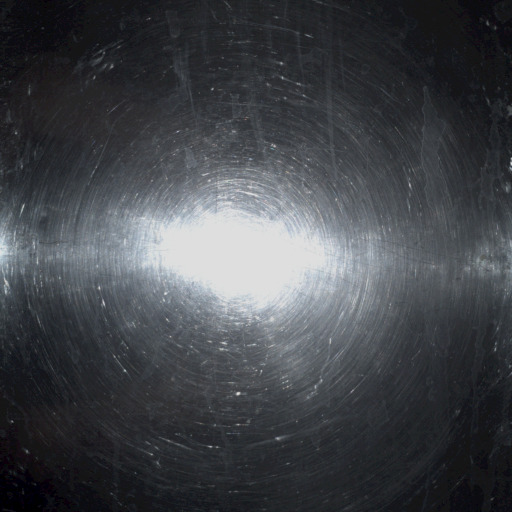} \\

    Ours Rerender & 
    \includegraphics[width=0.15\textwidth]{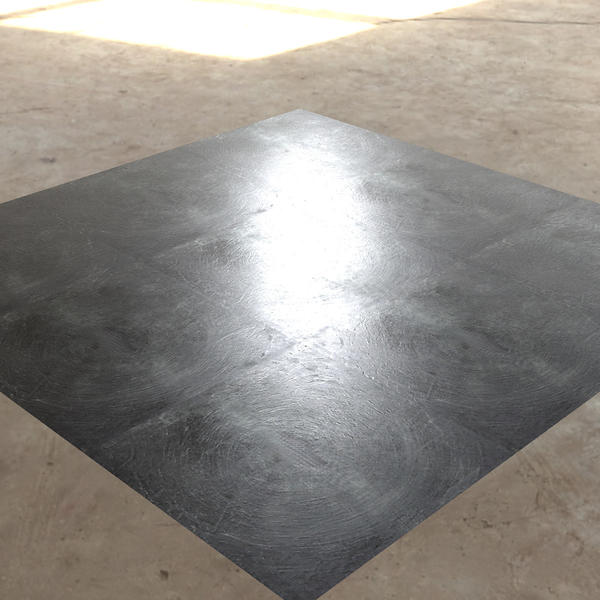} &
    \includegraphics[width=0.15\textwidth]{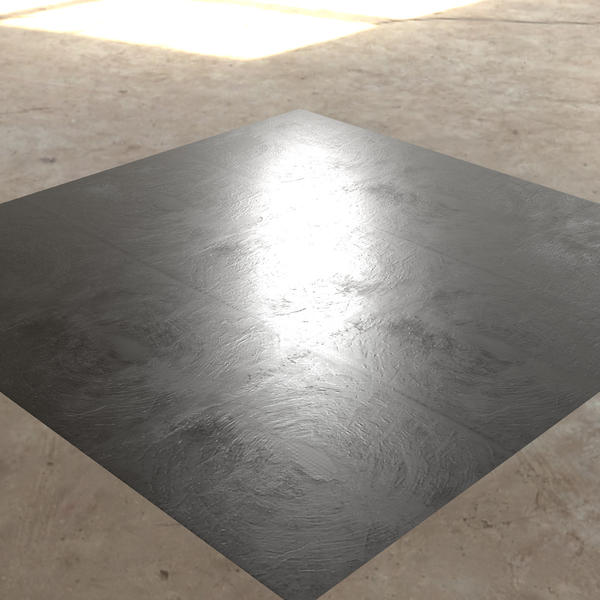} &
    \includegraphics[width=0.15\textwidth]{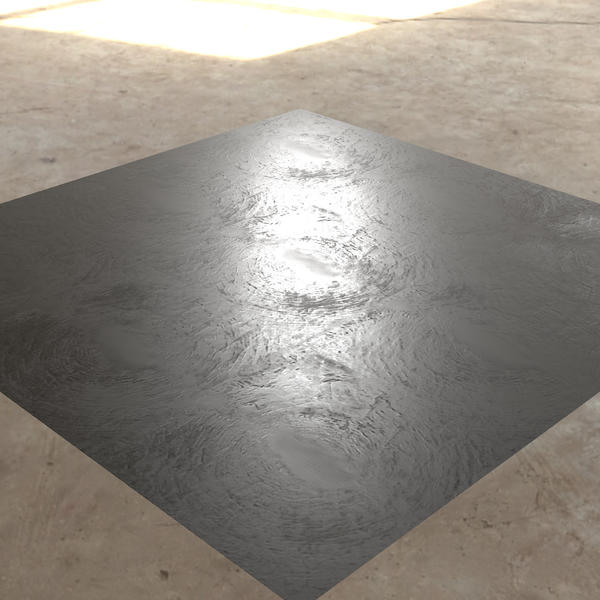} &
    \includegraphics[width=0.15\textwidth]{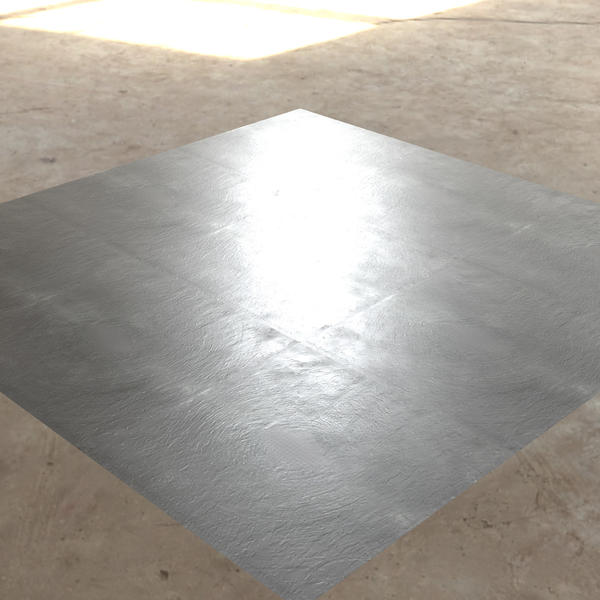} \\

\end{tabular}

    \caption{Capturing the same material with four different mobile phones leads to a rather consistent reconstruction with our approach \ie same shape of highlights and color. Please note how homogeneous our tiled surface is, as one would expect it.}
    \label{fig:comparison_phones}
\end{figure*}

In addition, we performed reconstructions of the same material with different mobile phones. While our approach results in very similar reconstructions independent of the brand, the approach by \citet{Deschaintre2018} is rather inconsistent (see Figure~\ref{fig:comparison_phones}).

\subsection{Limitations}

The proposed method provides generally good results. However, when the constraints are violated the results may differ. These constraints are based on violations of the capturing process and the scanned materials. For example, anisotropic materials are not reconstructable, because of the isotropic GGX term used in the Cook Torrance model. Other specific materials such as some textiles are also not well covered by this reflectance model, as a sheen is required to capture the microfibers of some cloth fully. Due to the single shot approach reconstruction of the exact Fresnel effect is not possible either. However, for casual material acquisition, the results are convincing.


By violating the capturing assumptions, the results of the predictions may degrade. If the distance to the material is drastically altered from the recommended 50cm, the scale of individual features differs. For example, the normal intensity may grow too strong if the material was captured from a close distance. If the distance is too far, the flash may hardly be visible, and thus the network is unable to recover the parameters accurately even though the training process operated with varying intensity of the flash. In the same vein, if the material is illuminated by other intense light sources, which overpower the camera flash, the prediction is not reliable either. Furthermore, in strongly reflective materials like car paints in brighter environments the reflection of the capturing device is visible. These reflections cannot be eliminated by the network, as the ambiguity of the reflections or darker areas on the surface cannot be resolved. For real-world capturing scenarios we highly recommend capturing RAW images, as our network is trained on linear data. Modern camera and mobile phones employ stark post-processing to provide visually pleasing results. The post-processing includes sharpening, saturation increase, shadowing lifting, smoother highlight roll off, brightness and contrast changes. All these adjustments are unknown to the network. When capturing RAW photographs, these adjustments are not applied automatically and thus do not violate our linear input data constraint.

\begin{figure}[htb!]
    \begin{center}
\begin{tabular}{c @{\hskip 0.07in} c @{\hskip 0.05in} c @{\hskip 0.05in} c @{\hskip 0.05in} c}

Input & Base color & Normal & Roughness & Metallic \\

\includegraphics[width=0.18\linewidth]{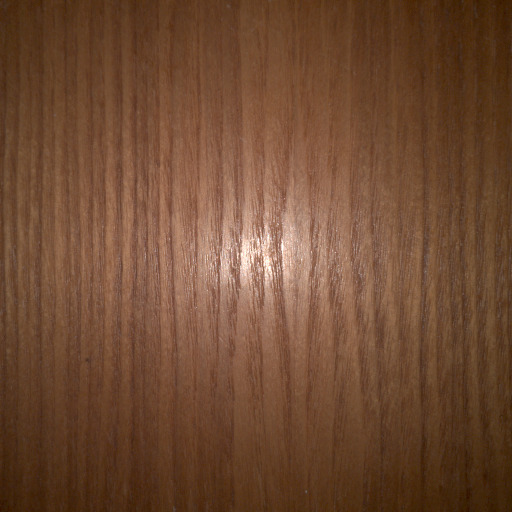} & 
\includegraphics[width=0.18\linewidth]{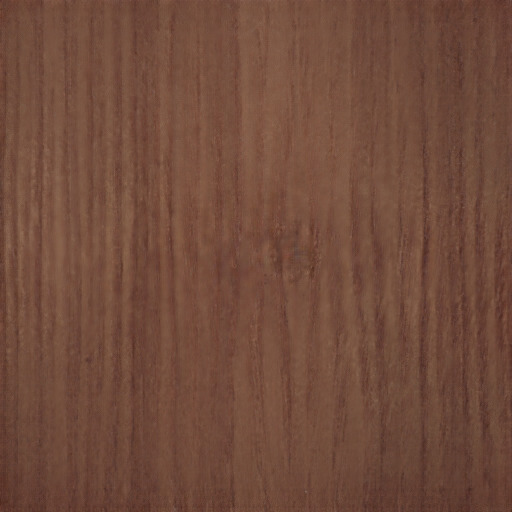} & 
\includegraphics[width=0.18\linewidth]{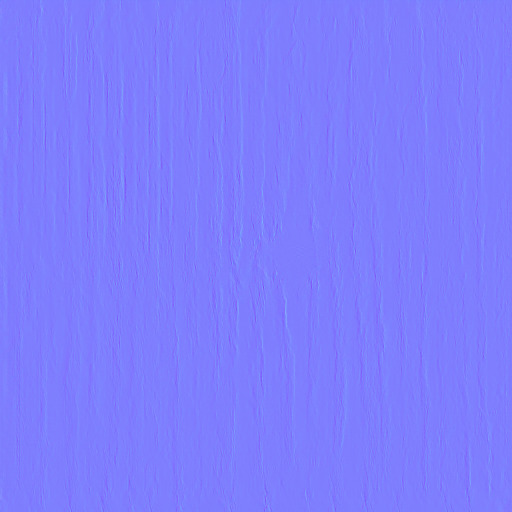} & 
\includegraphics[width=0.18\linewidth]{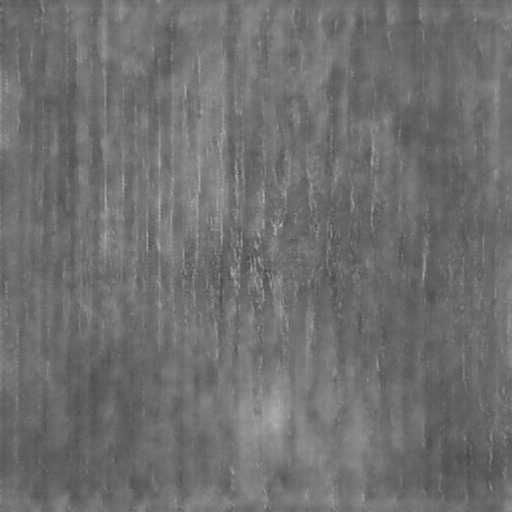} &
\includegraphics[width=0.18\linewidth]{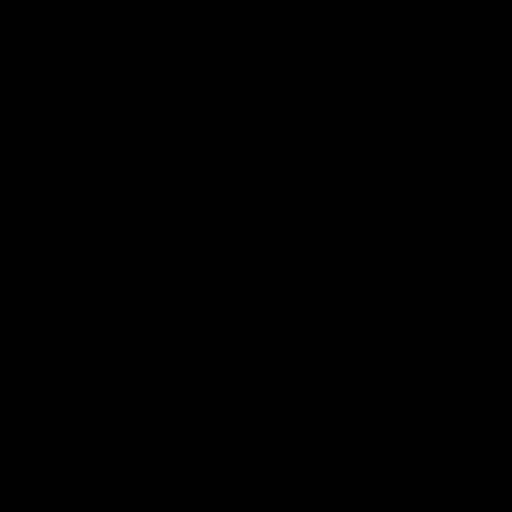} \\

\includegraphics[width=0.18\linewidth]{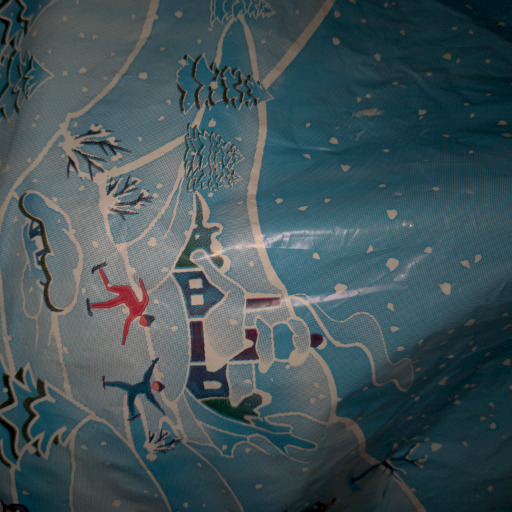} & 
\includegraphics[width=0.18\linewidth]{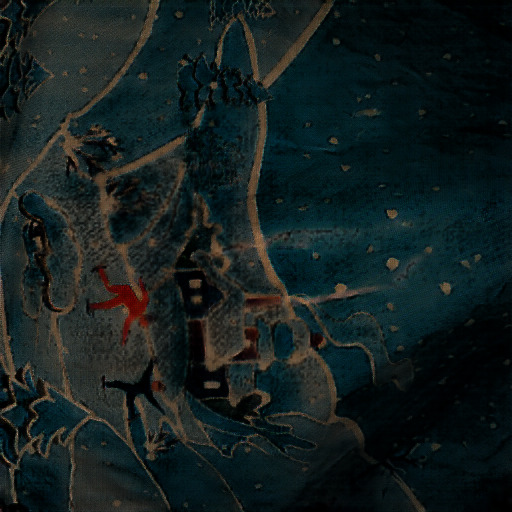} & 
\includegraphics[width=0.18\linewidth]{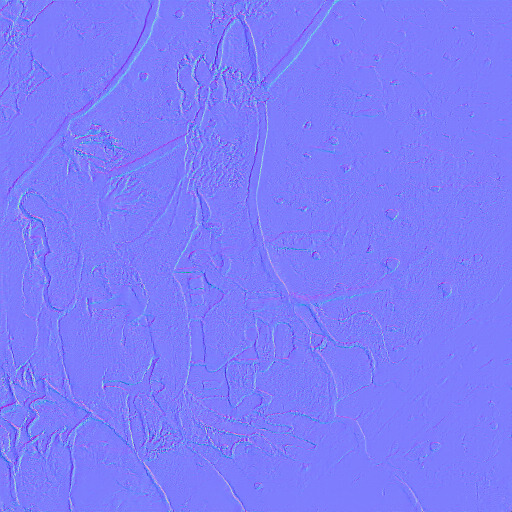} & 
\includegraphics[width=0.18\linewidth]{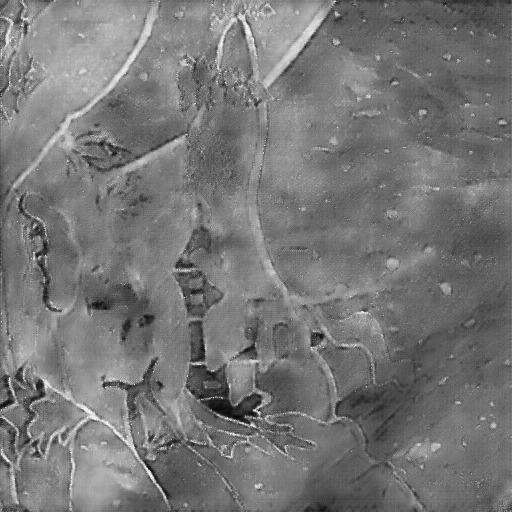} &
\includegraphics[width=0.18\linewidth]{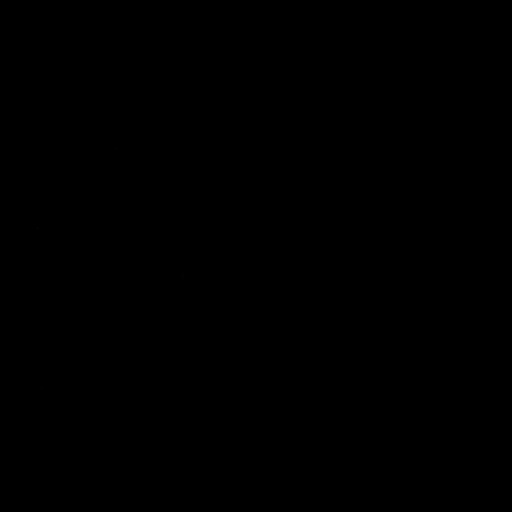} \\

\end{tabular}

    \end{center}
    \caption{The top row displays a positive effect of learned correlation. The normal and roughness detail are recovered plausibly. In the bottom row, the correlation leads to less favorable results. Here, the material texture introduce a complex structure to  the normal and roughness map.}
    \label{fig:correlation_example}
\end{figure}

One last violation which is on the other hand desirable in reproducing the human-made texture style is a correlation between parameters. For example in the top row of Figure~\ref{fig:correlation_example} a wood example is shown where the wood grain in the base color correlates with the normal and roughness map, and these maps can be reconstructed plausibly. However, in the bottom row of Figure~\ref{fig:correlation_example} this effect produces unrealistic results. The structure from the print is reproduced in the normal and roughness parameter maps.

\section{Future work}
It is shown that the method provides a reliable prediction on isotropic materials. However, many materials are anisotropic, and the learning process would need to consider both the strength and direction of the anisotropy. An increase in prediction resolution is equally desirable, as current games use textures in $4096\times4096$px resolution and upward. However, the GPU memory limitation prevented us in this work from increasing the resolution further. Introducing the popular perceptual loss based on a pre-trained VGG16 \cite{wang2018pix2pixHD, Johnson2016, enhancenet} could improve the prediction quality. Lastly, an extension to more general 3D geometry like \cite{Nam2018, Li2018a} while still maintaining a full Cook-Torrance model with both diffuse and specular lobes, is an interesting approach for future work.

\section{Conclusion}

We propose a framework for the reliable acquisition of SVBRDF from a single mobile phone flash image. By acknowledging the unconstrained capture environment of casual BRDF acquisition with environment map illuminated synthetic training data, the proposed method generalizes well to real data. Due to the introduction of a non-per-pixel loss based on a GAN approach, the resulting parameters capture the style of hand-authored materials better than previous work, while at the same time producing an accurate material reproduction with a believable specular behavior. Artifacts from secondary illumination and the harsh reflection of the camera flash are further reduced compared to previous work. At the same time, the increased resolution provides finer detail and the gap to actual usage of single image BRDF estimations in movie and video game productions is further reduced. It is demonstrated that this network can estimate a wide variety of different isotropic materials.

\paragraph{Acknowledgement}
Funded by the Deutsche Forschungsgemeinschaft (DFG, German Research Foundation) – Projektnummer 276693517 – SFB 1233

\bibliographystyle{ACM-Reference-Format}
\bibliography{ms}

\appendix

\section{Auto Exposure Calculation}
\label{sec:append_auto_exposure}

Our rendering algorithm produces High Dynamic Range (HDR) values. However, our network expects Low Dynamic Range (LDR) between 0 and 1. Therefore, the rendering output needs to be transformed without squashing the value range. To achieve this we calculate an ideal exposure multiplier which is applied to rendered result and the image can then be clipped to the value range of 0 to 1. Here, the values are first transformed to Exposure Values (EV). Per convention EV are defined for ISO 100 and are expressed as: $\text{EV}_{100} = \log_2 (\frac{N^2}{t})$, with $N$ being the aperture in f-stops and $t$ being the shutter time in seconds. different ISO speeds $S$ the EV can be changed with: $\text{EV}_S = \text{EV}_{100} + \log_2 \frac{S}{100}$. As $A$, $S$ and $t$ are unknown for the rendered images, this value can be calculated with the average scene luminance $L$ and the reflected-light meter calibration constant $K$: $\text{EV} = \log_2 \frac{L S}{K}$. A common value for $K$ is $12.5 \text{cd} \frac{s}{m^2} \text{ISO}$. The scene luminance is then determined by transforming the RGB HDR image to luminance and then calculating the average $L_\text{avg}$. This can then be used to calculate the photometric exposure $H=\frac{qt}{N^2}L=t E$, with $q$ being the lens and vignetting attenuation. Here, the Saturation Based Sensitivity (SBS) $H_\text{sbs}=\frac{78}{S_{\text{sbs}}}$ is used \cite{Kerr2007}. The maximum luminance is then defined as $L_\text{max} = \frac{78}{S} \frac{N^2}{q t}$. With a common value for $q$ of $0.65$ and the default ISO 100 for the Exposure Values, the maximum luminance is expressed as: $L_\text{max} = 1.2 * \text{EV}_{100}^2$. This $L_\text{max}$ value can now be multiplied with the RGB color image and clipped afterward to the LDR range. 

\end{document}